\documentclass[twocolumn,showpacs,prd,aps,tightenlines,superscriptaddress]{revtex4}

\usepackage{graphics,graphicx}
\usepackage{color}

\begin{document}

\title{Exploring black hole superkicks}

\author{Bernd Br\"ugmann, Jos\'e A. Gonz\'alez, Mark Hannam, Sascha Husa, Ulrich Sperhake}

\affiliation{Theoretical Physics Institute, University of Jena, 07743 Jena, Germany}

\begin{abstract}
Recent calculations of the recoil velocity in black-hole binary
mergers have found kick velocities of $\approx2500\,$km/s for
equal-mass binaries with anti-aligned initial spins in the orbital
plane. In general the dynamics of spinning black holes can be
extremely complicated and are difficult to analyze and understand. In
contrast, the ``superkick'' configuration is an example with a high
degree of symmetry that also exhibits exciting physics. We exploit the
simplicity of this test case to study more closely the role of
spin in black-hole recoil and find that: the recoil is with good
accuracy proportional to the difference between the $(l = 2, m = \pm
2)$ modes of $\Psi_4$, the major contribution to the recoil occurs
within $30M$ before and after the merger, and that this is after the
time at which a standard post-Newtonian treatment breaks down. We also
discuss consequences of the $(l = 2, m = \pm 2)$ asymmetry in the
gravitational wave signal for the angular dependence of the SNR and
the mismatch of the gravitational wave signals corresponding to the
north and south poles.
\end{abstract}

\pacs{
04.25.Dm, 
04.30.Db, 
95.30.Sf, 
98.80.Jk
}

\maketitle

\section{Introduction}

More than forty years after Hahn and Lindquist started the numerical investigation
of colliding black holes~\cite{Hahn64}, 
a series of breakthroughs starting in 2005 \cite{Pretorius:2005gq,Campanelli:2005dd,Baker:2005vv}
has turned the quest for stable black-hole inspiral simulations into a gold rush. 

A particular focus of the last few months
has been the so-called recoil or rocket
effect due to ``beamed'' emission of gravitational radiation
\cite{Bonnor1961,Peres1962,Bekenstein1973}.
By momentum conservation, radiation of energy in a preferred direction
corresponds to a loss of linear momentum and the black hole that results from
the merger thus recoils from the center-of-mass frame with speeds of
up to a few 
thousand km/s. The velocity of this ``kick'' depends on the configuration of the
system (e.g., the mass ratio and spins) and details of the merger dynamics,
but not on the total mass (velocity is dimensionless in geometric units).
From an astrophysical point of view, the recoil effect is particularly interesting
for massive black holes with masses $>10^5 \,M_{\odot}$, which exist at the
center of most galaxies and may have a substantial impact on the structure and
formation of their host galaxies. Observational consequences of large recoil have
recently been considered in \cite{Loeb:2007wz,Bonning:2007vt}.

The largest recoil effects have so far been found \cite{Gonzalez:2007hi,Campanelli:2007cg}
for a particularly simple configuration suggested in \cite{Campanelli:2007ew} based on
\cite{Kidder1995}: 
equal-mass binaries with (initially) anti-aligned spins in the orbital plane.
Based on numerical simulations for different configurations and a
post-Newtonian approximation~\cite{Kidder1995}, an estimate of
$1300~{\rm km/s}$ had been obtained for this configuration with
maximal spin~\cite{Campanelli2007v1}.
The kicks found in full numerical simulations are however even larger,
e.g.~$2500\,$km/s \cite{Gonzalez:2007hi} or $1800\,$km/s
\cite{Campanelli:2007ew,Campanelli:2007cg} for non-maximally
spinning black holes.
This is of the order of 1\% of the speed of light, and can be larger
than the escape velocity of about $2000\,$km/s from giant elliptical
galaxies.
Extrapolating current numerical results for non-maximal spins to
maximally spinning black holes predicts recoil velocities of up to
$\approx 4000$ km/s \cite{Campanelli:2007cg}.
Smaller but still significant kick
velocities have been found for several different types of black hole
configurations
\cite{Baker2006b,Herrmann2006,Gonzales06tr,Herrmann:2007ac,Koppitz2007,Campanelli2007,Tichy:2007hk,Herrmann:2007ex}. 
Estimations of the probabilities to obtain different kick velocities for
different mass ratios and high spins were studied in \cite{Schnittman:2007sn}.

The parameter space of the inspiral of spinning black holes is very large, 
and although its full exploration will require numerical methods, analytical understanding 
and approximations will be crucial to render the task economical.
The purpose of the present paper is to obtain a better understanding of the physics
that leads to the large kick results recently observed, and in particular to compare with
post-Newtonian approximations, and see where such approximations are accurate, and where
they (currently) break down.

We will refer to a configuration similar to that described in
\cite{Campanelli:2007cg,Gonzalez:2007hi}, i.e., two equal-mass black holes with spins
anti-aligned and in the orbital plane, as a superkick configuration.
The superkick configuration exhibits ``$\pi$ symmetry'',
i.e.~it is invariant under a rotation by an
angle $\pi$
about an axis perpendicular to the initial orbital plane.
It follows from this symmetry that linear 
momentum will not be
radiated in the $x$ or $y$ directions, but only in the $z$-direction.
As a consequence, the center-of-mass will remain at $(x=0,~y=0)$, but can
move in the $z$-direction.

The paper is organized as follows.
In Sec.\ \ref{sec:num} we briefly summarize our numerical methods,
and list the simulations we have performed.
Sec.\ \ref{sec:superkick_dynamics} analyses several aspects of the dynamics of the ``superkick''
configurations, in particular the comparison with post-Newtonian dynamics and various aspects
of the $(l = 2, m = \pm 2)$ asymmetry.  
Consequences of
this asymmetry for the angular dependence of the SNR
and the mismatch of the gravitational wave signals, exemplified by the extreme case
of the north and south poles, are discussed in Sec.\ \ref{sec:FF}.
The paper concludes with a discussion section and four appendices 
that contain post-Newtonian equations we use in this paper, and a number of
small results concerning the dynamics of moving-puncture simulations.

\section{Numerical methods and summary of simulations}\label{sec:num}

In this section we will summarize our numerical methods for evolving
black-hole binary spacetimes (largely by directing the reader to the relevant 
references), and specify the numerical simulations we
performed. The various simulations will be motivated more fully later in the
paper; for now we give an overview for later reference. 

We performed numerical simulations with the BAM \cite{Bruegmann:2006at,Bruegmann:2003aw} and LEAN 
\cite{Sperhake2006} codes, with modifications discussed in
\cite{Gonzalez:2007hi}. Both codes start with black-hole binary puncture
initial data \cite{Brandt97b,Bowen80} generated using a  
pseudo-spectral code \cite{Ansorg:2004ds}, and evolve them with the
$\chi$-variant of the moving-puncture \cite{Campanelli:2005dd,Baker:2005vv} version
of the BSSN \cite{Shibata95,Baumgarte99} formulation of the 3+1 Einstein
evolution equations \cite{York79}. The gravitational waves emitted by
the binary are calculated from the Newman-Penrose scalar $\Psi_4$, and the
details of this procedure for BAM and LEAN are given in \cite{Bruegmann:2006at}
and \cite{Sperhake2006}, respectively.

The parameters of our simulations are summarized in 
Table~\ref{tab:parameters}. Each black hole has mass $M_i$ 
(with mass parameter $m_i$ in the puncture data construction \cite{Brandt97b}), 
and the total 
mass is $M = M_1 + M_2$. The black holes have a coordinate separation of $D$. In
all runs the punctures are placed on the $y$-axis at $y = \pm D/2$ and given
momenta $p_x$ and spins $S = 0.723 M_i^2 = 0.2$. The spins are
aligned with the $y$-direction, except for the runs in the
``$\alpha$-series'', which are characterized by 
$S_y = \pm S \cos \alpha$ and $S_x = \mp S \sin \alpha$. 

The $\alpha$-series and $P$-series simulations used modifications of the MI configuration
described in \cite{Gonzalez:2007hi}. This configuration was chosen because the results
showed clean fourth-order convergence and high accuracy. We have found that
the resolution requirements increase 
significantly for simulations of spinning
black holes, and the MI configuration, with a small initial separation and
therefore short evolution time, provided a convenient starting point for our
study; these simulations also capture most of the important dynamics that we
wish to study.

\begin{figure}[ht]
\includegraphics[width=8cm]{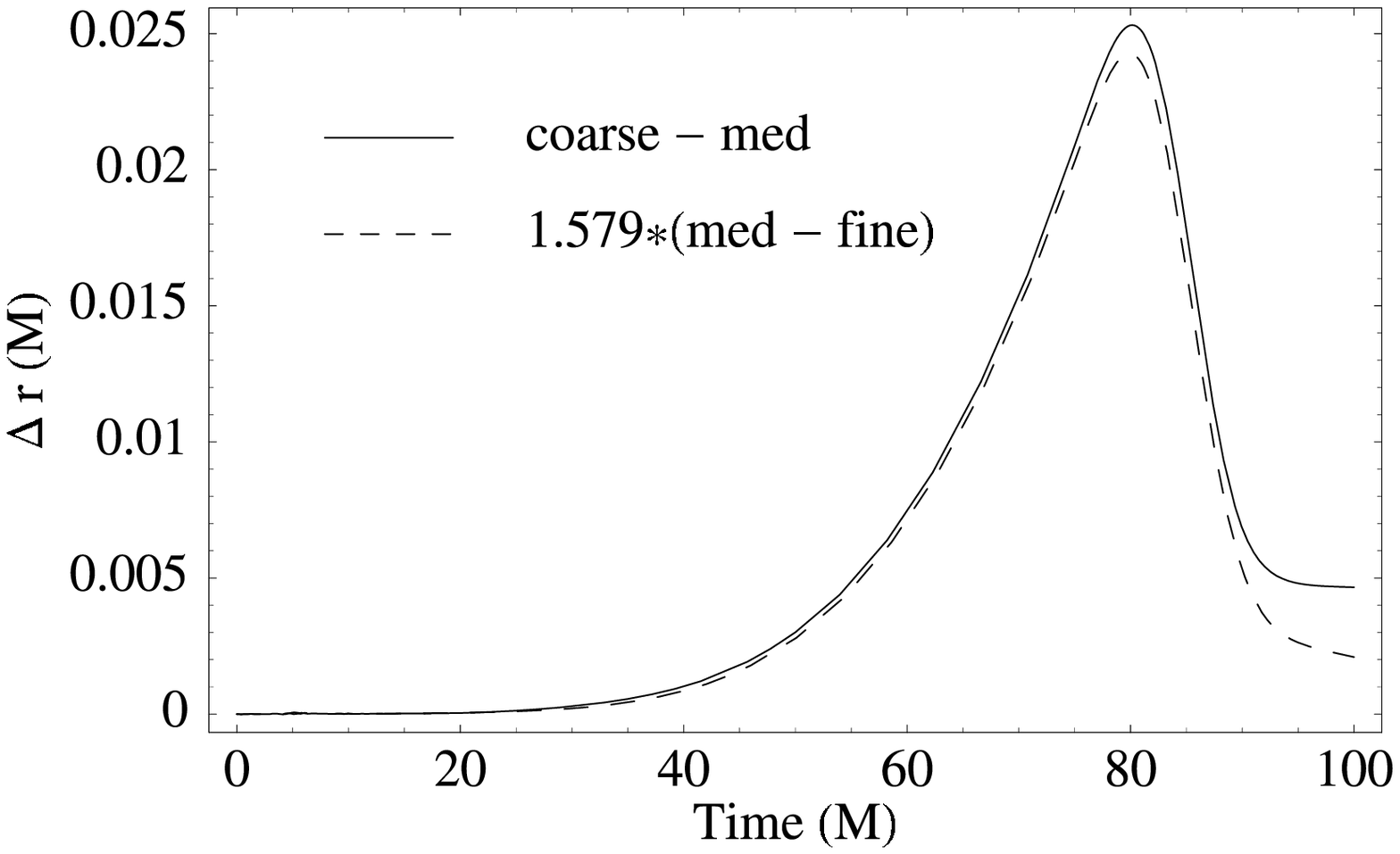}
\includegraphics[width=8cm]{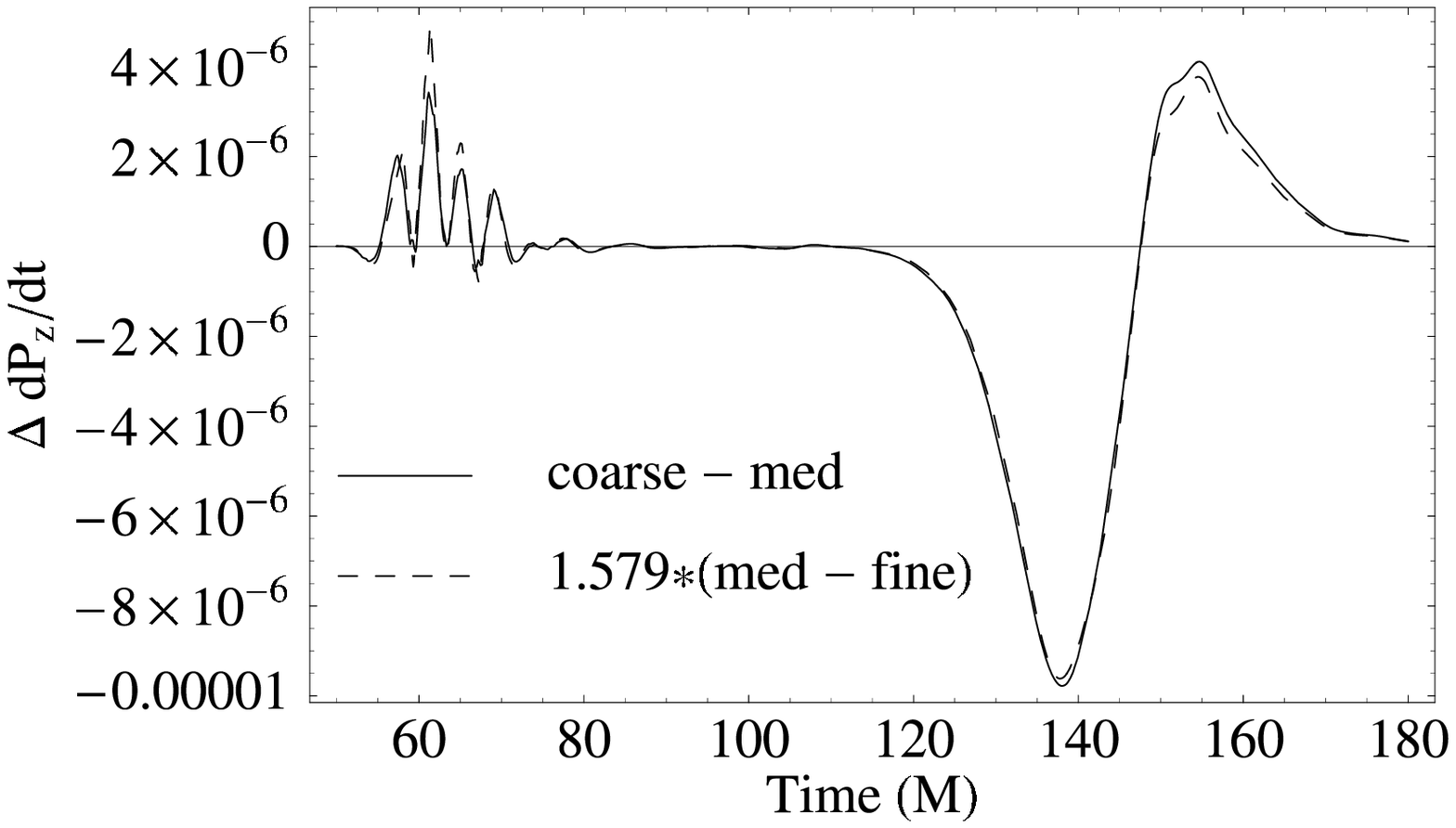}
\caption{\label{fig:AlphaConv} 
Convergence plots for the puncture separation $r$ and the linear momentum
radiation in the $z$-direction, $dP_z/dt$ obtained for model $\alpha=0$
of the $\alpha$-series. The plots are scaled consistent
with fourth-order convergence. 
After merger at about $t = 88M$ convergence in the puncture separation
is lost (as expected).
}
\end{figure}

The $\alpha$- and $P$-series simulations were performed with the grid setup
$\chi_{\eta=2}[6\times 44:4\times  88:6][88:5.82]$ in the notation of
\cite{Bruegmann:2006at}, i.e., the six inner boxes had $44^3$ points, the four 
outer boxes had $88^3$ points, the resolution on the finest level is $M/88$, and
the resolution at the outer boundary is $5.82M$. Convergence tests were
performed for the $\alpha = 0$ case (which is the same as the MI configuration
in \cite{Gonzalez:2007hi}) with inner-box sizes of $40, 44, 48$, and
corresponding resolutions. Clean fourth-order convergence of the linear
momentum radiation flux $dP_z/dt$ is shown in Figure~\ref{fig:AlphaConv}.
Also shown is convergence in the puncture separation, which is not
expected to last beyond the merger time of about $t=88M$ since the
separation between the two punctures inside the common apparent horizon
quickly approaches zero \cite{Bruegmann:2006at}.

Further simulations were performed with larger initial separation and
with quasi-circular orbit parameters (calculated according the
prescription given in Appendix~\ref{app:qc}). These are indicated D6
(for $D = 6M$) and D8 (for $D = 8.2M$) in the table. 
The D8 simulation
was performed using the LEAN code, while all others were performed
with BAM. The grid setup for the D6 simulations was the same as for
the $\alpha$- and $P$-series, and the convergence test referred to
later used inner box sizes of 44, 48 and 52 points.  The D8 simulation
used a grid setup 
$\chi_{\eta=1}\left[2\times133:
1\times155:2\times133:3\times67:9
\right]\left[ 44:\frac{32}{11} \right]$, 
where the innermost three levels with 67 points are centered around
either hole and follow the motion of the puncture.

\begin{table}
\begin{ruledtabular}
\begin{tabular}{||l|r|r|r|r|r|}
\hline
Simulation & $D$ & $m_i$ & $p_x$  & $M$ & $\alpha$ \\
\hline 
$\alpha$-series & 6.514 & 0.363 & 0.133    & $1.052$     & $0 \leq \alpha < 2\pi$ \\    
                              &            &             &                &                     &  $\delta \alpha = \pi/6$ \\
 \hline  
$P$-series          & 6.514 & 0.363 & $0.13034 \leq p_x$ & 1.052 & 0 \\
                              &            &             & $\leq 0.13566$ &        &    \\     
                              &            &             & $\delta p = 0.003325$ &              &    \\
\hline
D6                & 6.0       & 0.296      & 0.1382    &  $1.0$     & 0 \\
\hline
D8                & 8.198   & 0.2875   & 0.11          & $1.0$        & 0 \\
\hline
\end{tabular}
\end{ruledtabular}
\caption{
\label{tab:parameters}
Physical parameters of the simulations performed for this paper. 
}
\end{table}

Experimentally we have observed that the resolutions used in the
$\alpha$- and $P$-series simulations are not sufficient to obtain
clean convergence for evolutions of spinning black holes orbiting for
longer periods of time.
It thus appears that
the good convergence results for these particular series are
largely due to
the close initial separation of the black holes, which results in a rather
quick merger time of about $88M$. When the black holes are placed further
apart (or even making the seemingly innocuous change of choosing quasicircular
orbit initial parameters for the same separation as the $\alpha$-series
simulations) convergence is lost before the black holes merge.
We expect that fourth-order convergence would be obtained if 
sufficiently high resolutions were used, but the extra computational 
expense was not necessary for the analysis in this paper. 

\begin{figure}[ht]
\includegraphics[width=8cm]{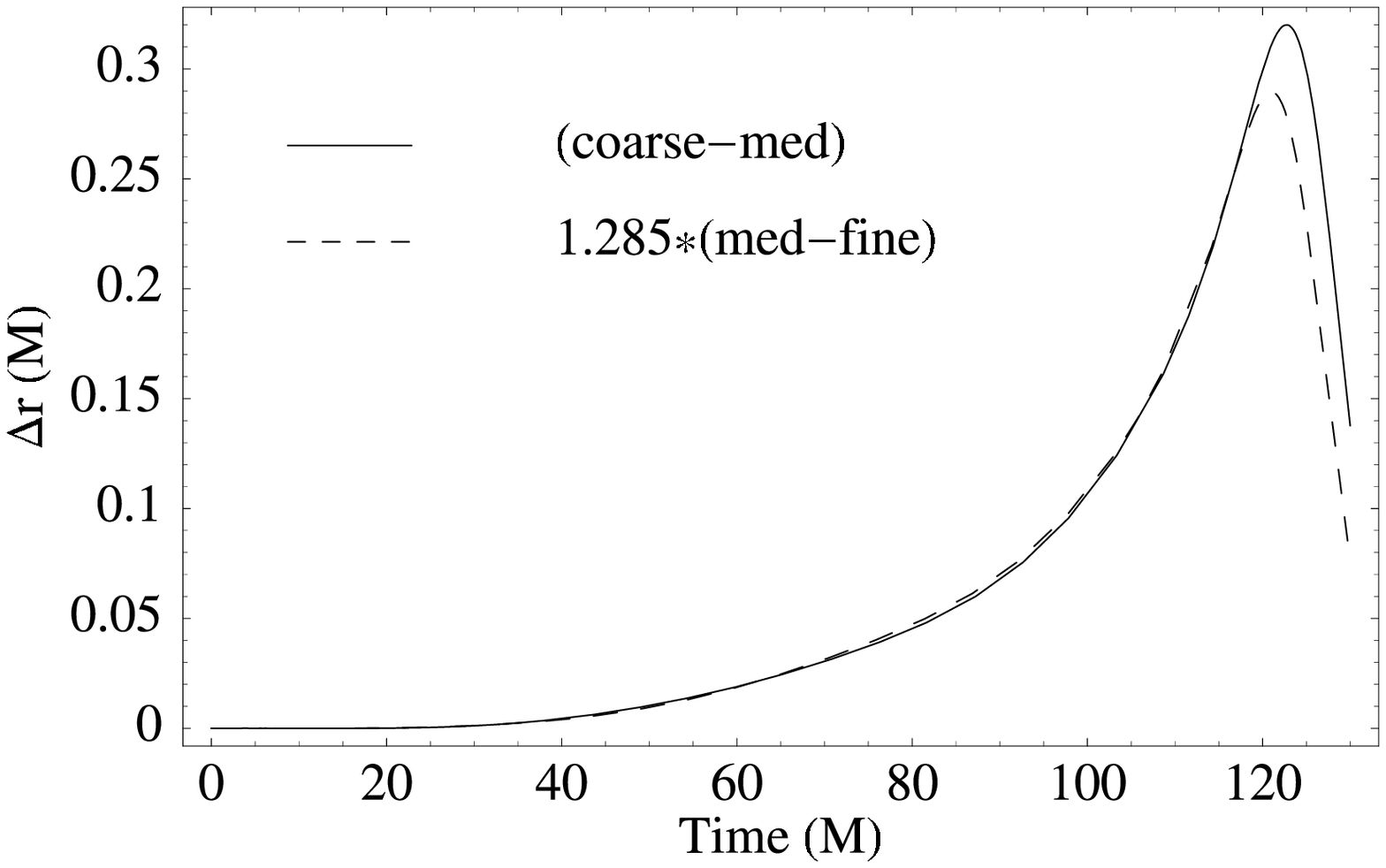}
\includegraphics[width=8cm]{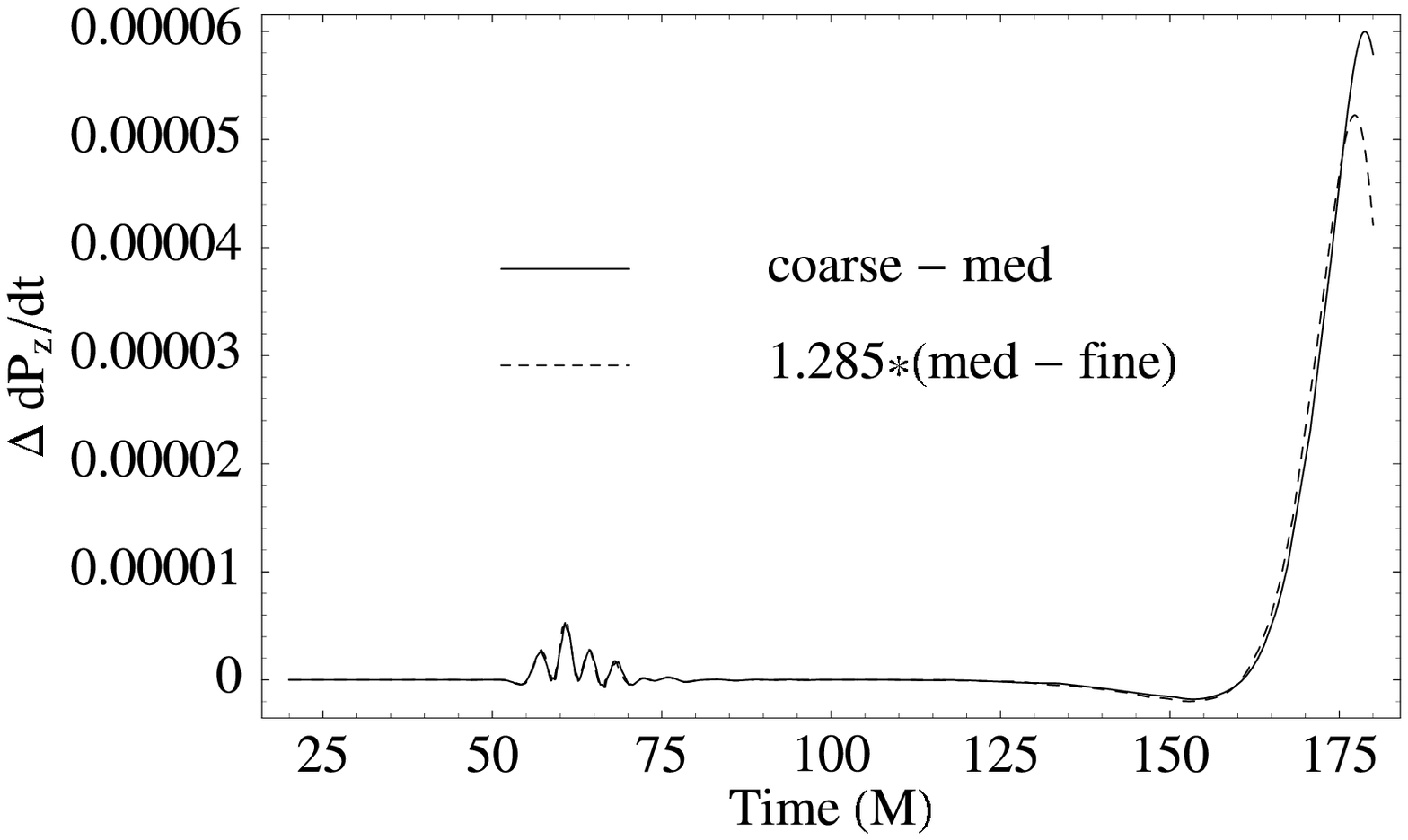}
\caption{\label{fig:R3conv} 
Convergence of the puncture separation and $dP_z/dt$ as functions of time for
evolutions of model D6. Results are
scaled for fourth-order convergence. We see that fourth-order convergence
is lost in the puncture separation  at about $t = 115M$, which corresponds
to roughly $t = 165M$ in quantities from waves extracted at $R_{ex} = 50M$,
which is about when we see a loss of convergence in $dP_z/dt$. Note that
we cut the plot at $t\approx 175~M$ when convergence is lost.
}
\end{figure}

\begin{figure}[ht]
\includegraphics[width=8cm]{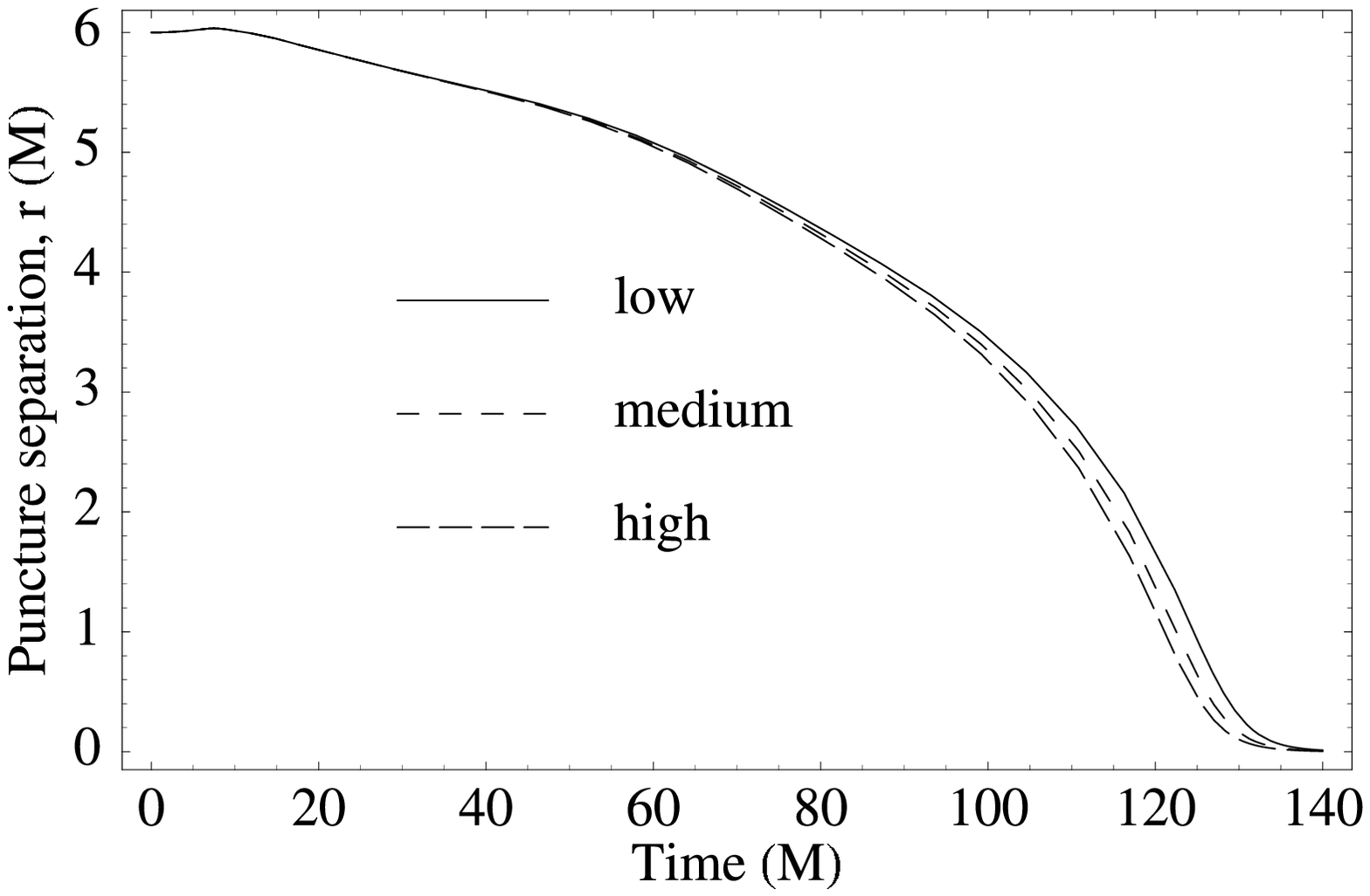}
\includegraphics[width=8cm]{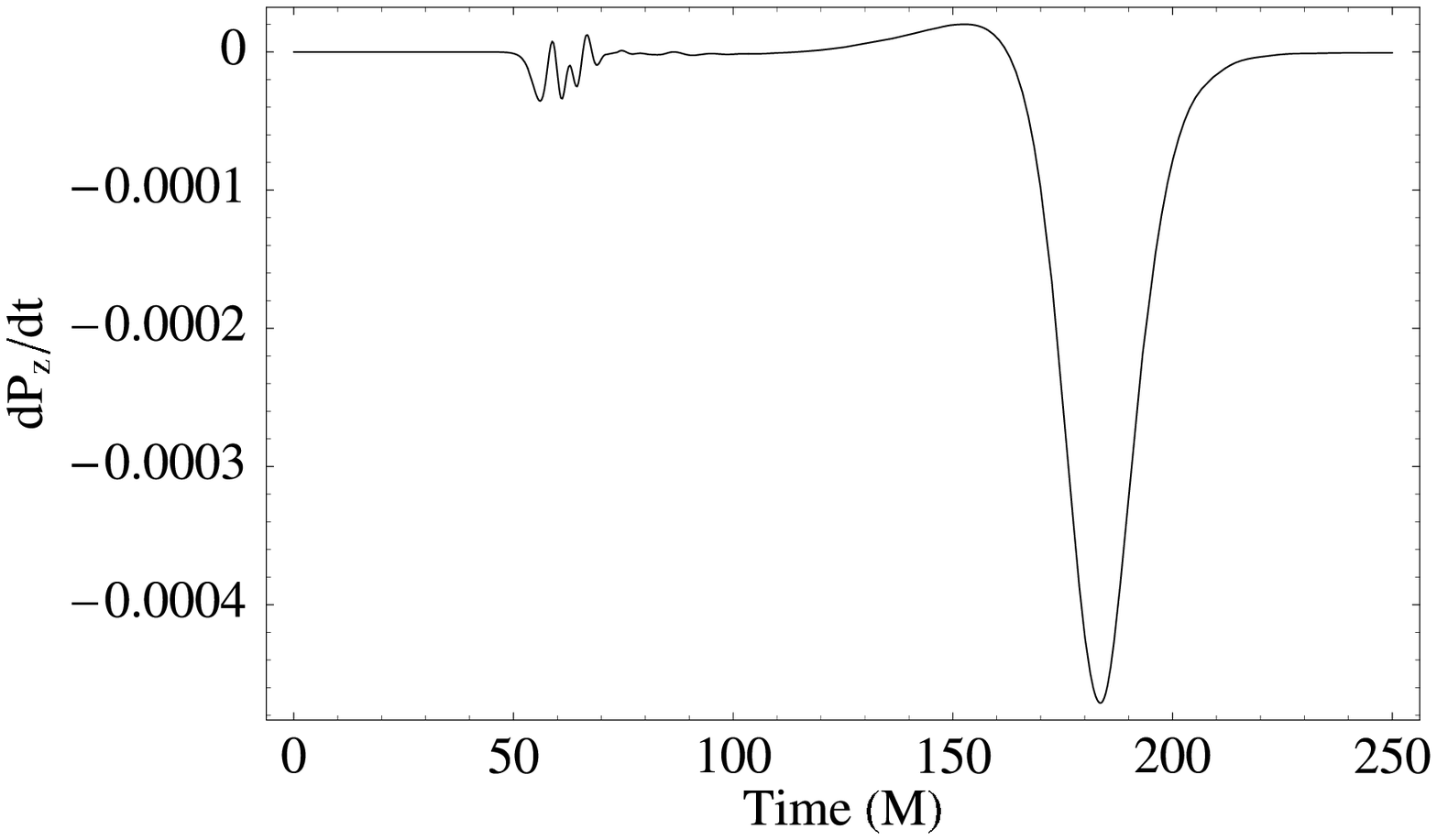}
\caption{\label{fig:R3motion} 
Puncture separation and $dP_z/dt$ as functions of time for the evolutions
of model D6. Results from low, medium and high
resolution simulations are shown. Only the highest resolution is shown for
$dP_z/dt$.
}
\end{figure}

In the D6 simulations we find that the puncture separation and linear
momentum radiation flux $dP_z/dt$ converge well for up to $15M$ before merger,
as shown in Figures~\ref{fig:R3conv} and \ref{fig:R3motion}. Note that since 
the waves are extracted at $R_{ex} = 50M$, we need to take into account a time
lag of roughly $50M$ when comparing times related to puncture motion and 
wave extraction. These simulations
will be used only for discussions of 
the qualitative behavior, and for analysis at early times, when we are
confident that the results are reliable. Similarly the long D8
LEAN simulation
will only be used for qualitative comparison with post-Newtonian
results. 

\section{Analysis of Superkick dynamics}\label{sec:superkick_dynamics}

We begin by describing the dynamics of two black holes in a superkick
configuration with varying degrees of simplicity, in order to build up a
clearer picture of the physics, and to motivate the simulations and analysis
we have performed. In the simplest picture we draw (Section~\ref{sec:newtonian}) the
spin decouples from the orbital dynamics; a more complex picture includes
spin precession effects (Section~\ref{sec:precession}), and considering the
dynamics in full general relativity (GR) in Section~\ref{sec:duration} allows us to study the merger regime,
from which most of the kick effect originates. The full GR results can then be directly
compared  with PN predictions, which we do in Section~\ref{sec:PNcomparison}.
We discuss the spin of the final black hole in Section~\ref{sec:finalspin}.

\subsection{Kick velocity and $l=2, m=\pm 2$ symmetry breaking}

As noted before, the superkick configuration exhibits ``$\pi$ symmetry'' ($\phi \rightarrow \phi
+ \pi$), thus linear momentum will not be
radiated in the $x$ or $y$ directions, but radiation of linear momentum in the $z$
direction is allowed, and the center-of-mass will only
move in the $z$-direction.

As in nonspinning equal-mass binary simulations, almost all of the energy is
radiated in the $l=2,m=\pm 2$ modes: the maximal relative deviation of the
energy in those modes from the total energies
is roughly 2 \%, neglecting the contribution
from the junk radiation. This fact, and the symmetry discussed in the 
preceeding paragraph, leads us to expect that we should be able to directly
relate the kick in the $z$-direction to the imbalance between the $m=2$
and $m=-2$ modes, i.e., the difference in energy that is radiated toward
the ``north'' and ``south'' hemispheres. Using the special relativistic
relation $\vert \vec p \vert = E$ between the momentum and energy of a wave
packet (traveling at the speed of light), we expect a relation
\begin{equation}\label{eq:p_asym}
p_z = f\times (E_{22} - E_{2-2})
\end{equation} 
for the radiated momentum in the $z$-direction,
where $E_{22}$ and
$E_{2-2}$ are the energies radiated in the  $l=2,m=\pm 2$ modes, and $f$ is a
geometric factor.
Here $0 \leq f < 1$ expresses the fact that the radiation is smeared out in solid angle 
rather than sharply peaked in the direction of the poles.
Neglecting all modes but $l=2,m=\pm 2$,
we assume a wave signal in the form $\Psi_4 = \kappa F(t)
Y^{-2}_{22}(\theta,\phi) + \lambda \bar{F}(t) Y^{-2}_{2-2}(\theta,\phi) $, where
$\kappa,\lambda$ are real numbers ($\kappa=\lambda$ in the nonspinning equal mass case), 
$F(t)$ is a complex time dependent
function, and the $Y^{-2}_{2\pm2}$ are the spin-weighted spherical harmonics 
\begin{eqnarray}
  Y^{-2}_{2-2} &=& \sqrt{\frac{5}{64\pi}} \left( 1 -\cos \theta \right)^2
       e^{-2i\phi}, \nonumber \\
  Y^{-2}_{22} &=& \sqrt{\frac{5}{64\pi}} \left( 1 +\cos \theta \right)^2
       e^{2i\phi}.
\end{eqnarray}

\begin{figure}[ht]
  \includegraphics[width=8cm]{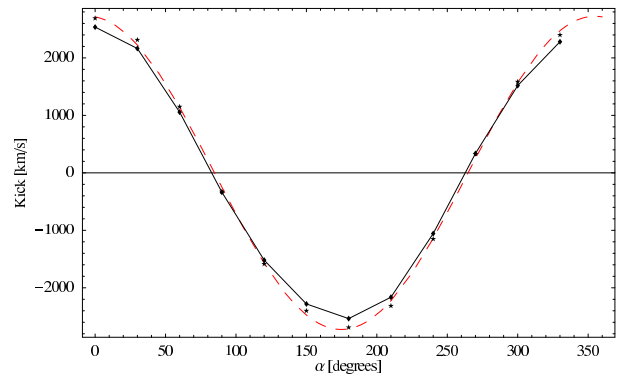}
  \caption{\label{fig:kick_vs_DeltaE}
           Comparison of the kick velocity in km/s according
           to Eq.\ (\ref{eq:p_asym}),
           for a range of angles $\alpha$.
           Data points for the measured kick and the estimate Eq.\ (\ref{eq:p_asym}) are
           shown, the points corresponding to the energy differences
           are connected. An analytical fit, $v_z = 2725 \cos(176+\alpha)$, to the measured kick is shown as a
           dashed line. Note that Eq.\ (\ref{eq:p_asym}) slightly underestimates the kick,
           which is consistent since it neglects contributions
           from higher order multipoles $l>2$.
           }
\end{figure}

Inserting this ansatz into the expressions for radiated energy and 
linear momentum (see e.g.\ Eqs.\ (48) and (49) in \cite{Bruegmann:2006at})
we obtain
\begin{eqnarray}
  \frac{dE}{dt} &=& \frac{r^2}{16\pi} \left( \kappa^2 + \lambda^2 \right)
  \left| \int_{\infty}^{t} F(\tilde t) d\tilde t \right|^2\,, \\
  \frac{dP_z}{dt} &=& \frac{2}{3}\frac{r^2}{16\pi} \left( \kappa^2 - \lambda^2
  \right) \left| \int_{\infty}^{t} F(\tilde t) d\tilde t
  \right|^2.
\end{eqnarray}
Consequently the value of the geometric factor $f$ can be determined as $f = 2/3$.
We find this relationship satisfied to very good accuracy in our numerical evolutions,
as shown in Figure~(\ref{fig:kick_vs_DeltaE}).

\begin{figure}[ht]
\includegraphics[width=9cm]{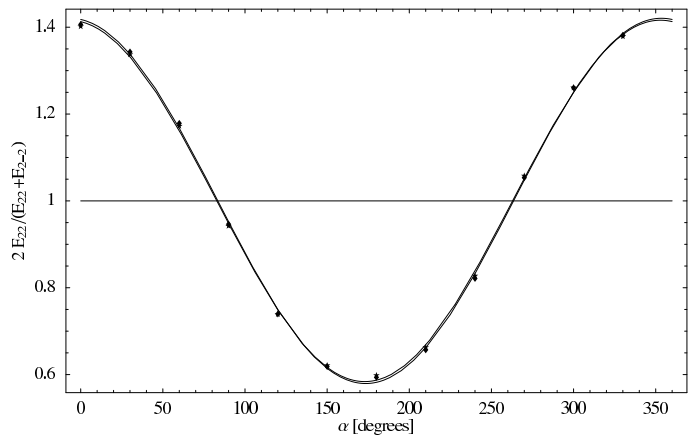}
\caption{\label{fig:DeltaE}
         Excess energy in the $l=2,m=2$ mode, ${2 E_{22}}/{(E_{22}+E_{2-2})}$
         plotted for extraction radii $R_{ext} = 30~M$, and $50~M$.
         The curves are the analytical fits for both extraction radii, see
         Eq.\ (\ref{eq:excess_fit})
         Clearly, there is no significant dependence of this ratio
         on extraction radius.}
\end{figure}

The relative asymmetry in the energies emitted in the $l=2,m=\pm 2$
modes, $2 E_{22}/(E_{22}+E_{2-2})$ (this quantity is unity when
there is no symmetry breaking) is plotted in Fig.~\ref{fig:DeltaE},
showing a maximal excess of roughly 40\%.
An analytic fit for extraction radius $ R_{ext} = 50 M $ is 
\begin{equation}\label{eq:excess_fit}
\frac{2 E_{22}}{(E_{22}+E_{2-2})} = 1 + 0.416 \cos({0.125 + \alpha}) \, .
\end{equation}
In the extreme case this fit corresponds to $E_{22}/E_{2-2} \approx 2.4$.
For this fit the statistical error from 95 \% confidence in the phase is roughly 20\%,
and roughly 2\% for the amplitude of the oscillation. Fits corresponding to the extraction radii
$R_{ext} = 30M$ and $75M$ give
consistent results with Eq.~(\ref{eq:excess_fit}) within the statistical error bars.

\subsection{Simplest assumption: spin decouples from black-hole dynamics}
\label{sec:newtonian}

As a first 
approximation of the dynamics, we may imagine that the
black holes behave like (force-free) gyroscopes in flat spacetime,
and as they orbit each other their
individual spin vectors do not change direction. For example, if the
spins were originally $\mathbf{S}_1 = (0,S,0)$ and $\mathbf{S}_2 = (0,-S,0)$,
these vectors would be constant throughout the evolution. If we also assume
that the spins do not noticeably influence the motion in the $x$-$y$ plane, then
simulations that start with the same initial separation and momenta will
display the same dynamics no matter how the spins are directed in the orbital
plane. This is the situation at the first-PN
approximation, since spin-orbit and spin-spin couplings enter at higher order.

\begin{figure}[ht]
\includegraphics[width=9cm,height=3.5cm]{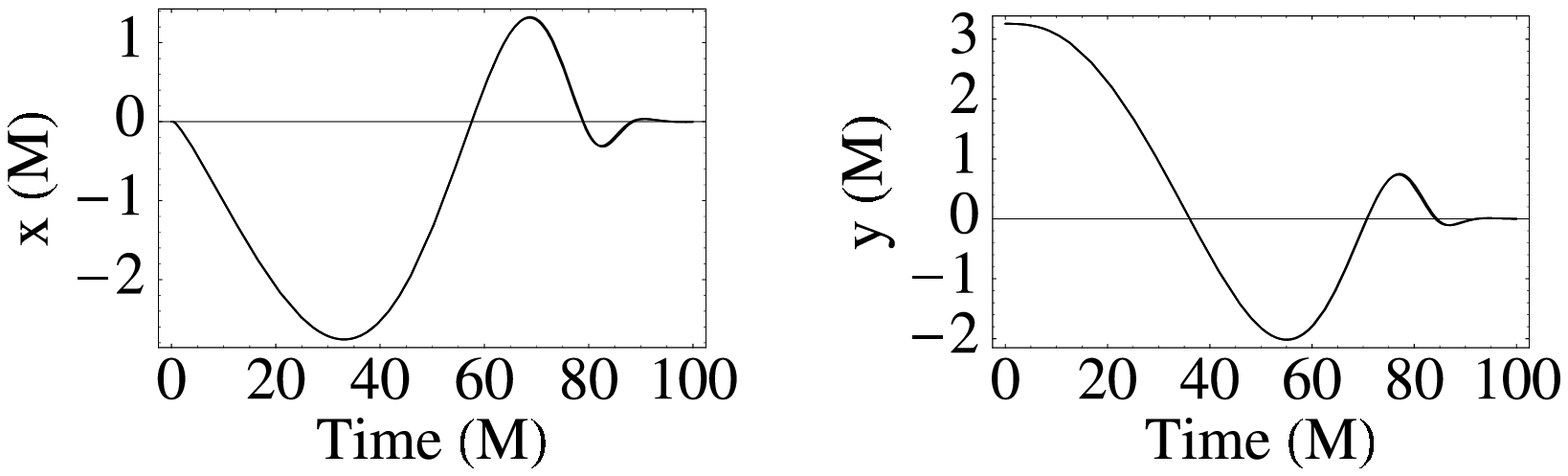}
\includegraphics[width=9cm,height=3.5cm]{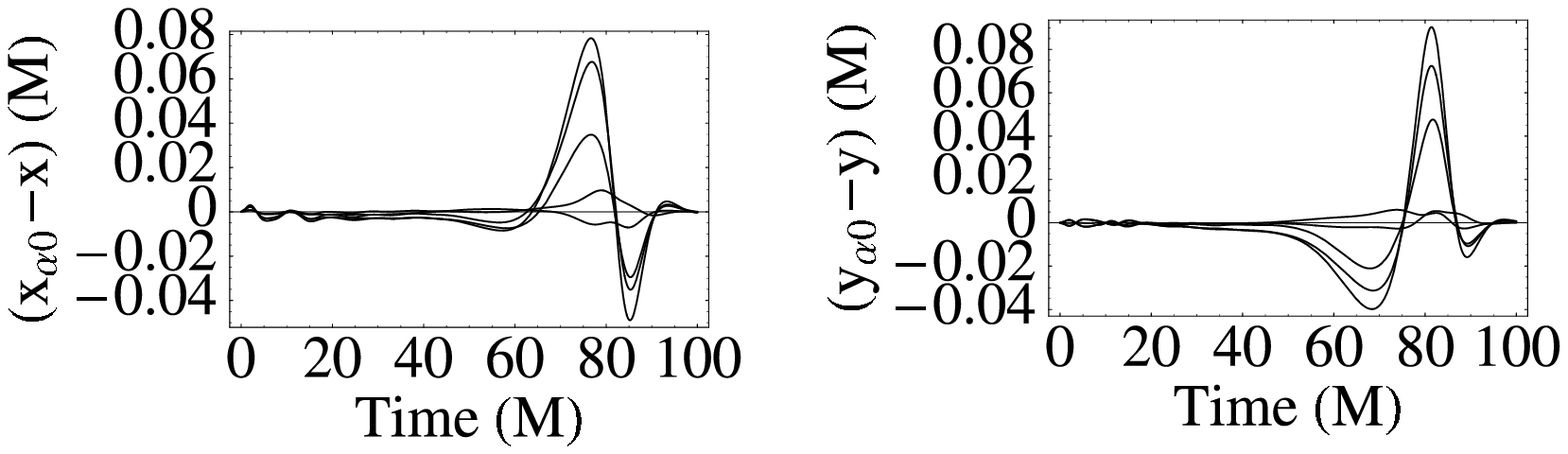}
\caption{\label{fig:alphamotion} 
The $x$- and $y$-motion of one of the punctures in six simulations from the
$\alpha$-series. The dynamics in the $x$-$y$ plane are almost identical for all
of the simulations (only one curve is actually visible in the upper
panels). The small variations in the motion (measured with respect
to the $\alpha = 0$ simulation), are shown in the lower panels.}
\end{figure}

This picture is surprisingly close to the observed dynamics in
numerical simulations. Figure~\ref{fig:alphamotion} shows the orbital
motion in the first six simulations of the $\alpha$-series, each
differing only in the initial directions of the spins (i.e.,
$\mathbf{S}_{1,2} = S (\mp \sin \alpha, \pm \cos \alpha, 0)$ and
$\alpha = 0...\pi$ in steps of $\pi/6$).
The motion shows differences as $\alpha$ is varied (shown in the lower
panels of the figure), but these differences are very small. In contrast, the
resulting kick from these simulations, shown in Figure~\ref{fig:kick_vs_DeltaE}, displays a clear
sinusoidal dependence on the angle --- the kick varies from $\approx -2500$
km/s to $+ 2500$ km/s. (Note that, since these values were calculated at a
small radiation extraction radius $R_{ex} = 50M$, the values in the plot are systematically
higher than the correct values by about 10\%). Similar figures were also shown
in \cite{Campanelli:2007cg}. 

We might conclude from these results that the final kick depends only on the
initial spin magnitude and direction of each black hole. One may write an
expression similar to Eq.\ (1) in \cite{Campanelli:2007cg}, which for the
superkick case reduces to \begin{equation}
V_K = k \cos (\alpha - \alpha_0), \label{eqn:trigkick}
\end{equation} and determine the constants $k$ and $\alpha_0$, which are
approximately given by $k \approx 2500~{\rm km/s}$
and $\alpha_0 \approx 0$ for our data. If this simple
picture was correct, and the spins really did behave as gyroscopes
in flat spacetime, then (\ref{eqn:trigkick}) would
allow us to determine $\alpha_0 = 0$ as the spin direction that produces
the maximum kick.

\subsection{The effect of spin precession during inspiral}
\label{sec:precession}

Needless to say, the real situation is more complicated. The picture
of the black holes' spins as gyroscopes in flat spacetime is valid
only at 1PN order; at higher post-Newtonian orders and in full
GR the spin directions evolve during the inspiral.
We expect that the magnitude of the kick depends on the magnitude and
direction of the spins when the black holes are close to merger. The
spin configuration at merger time is a function of the initial
configuration plus precession effects during the evolution.

\begin{figure}[ht]
\includegraphics[width=8cm]{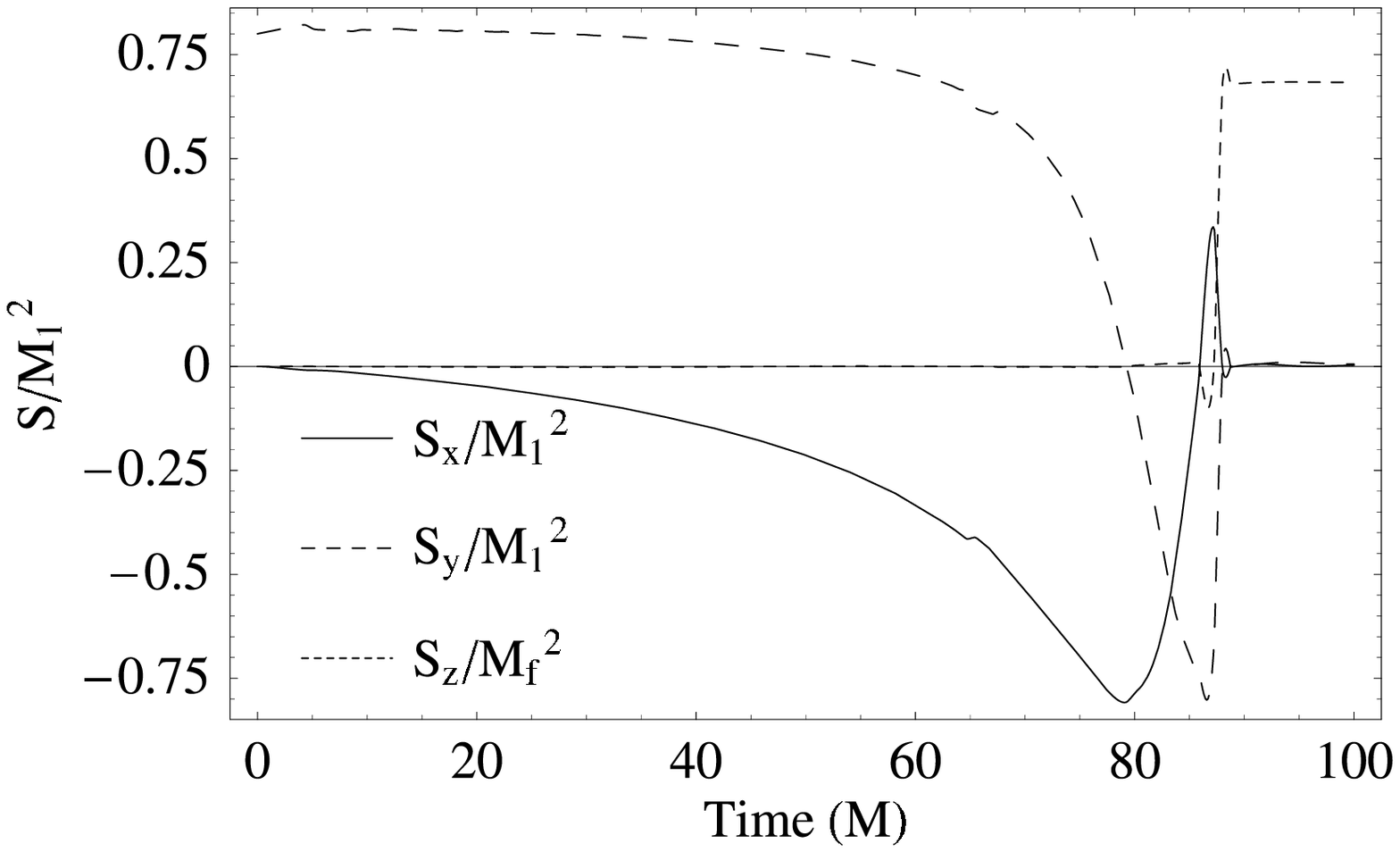}
\includegraphics[width=8cm]{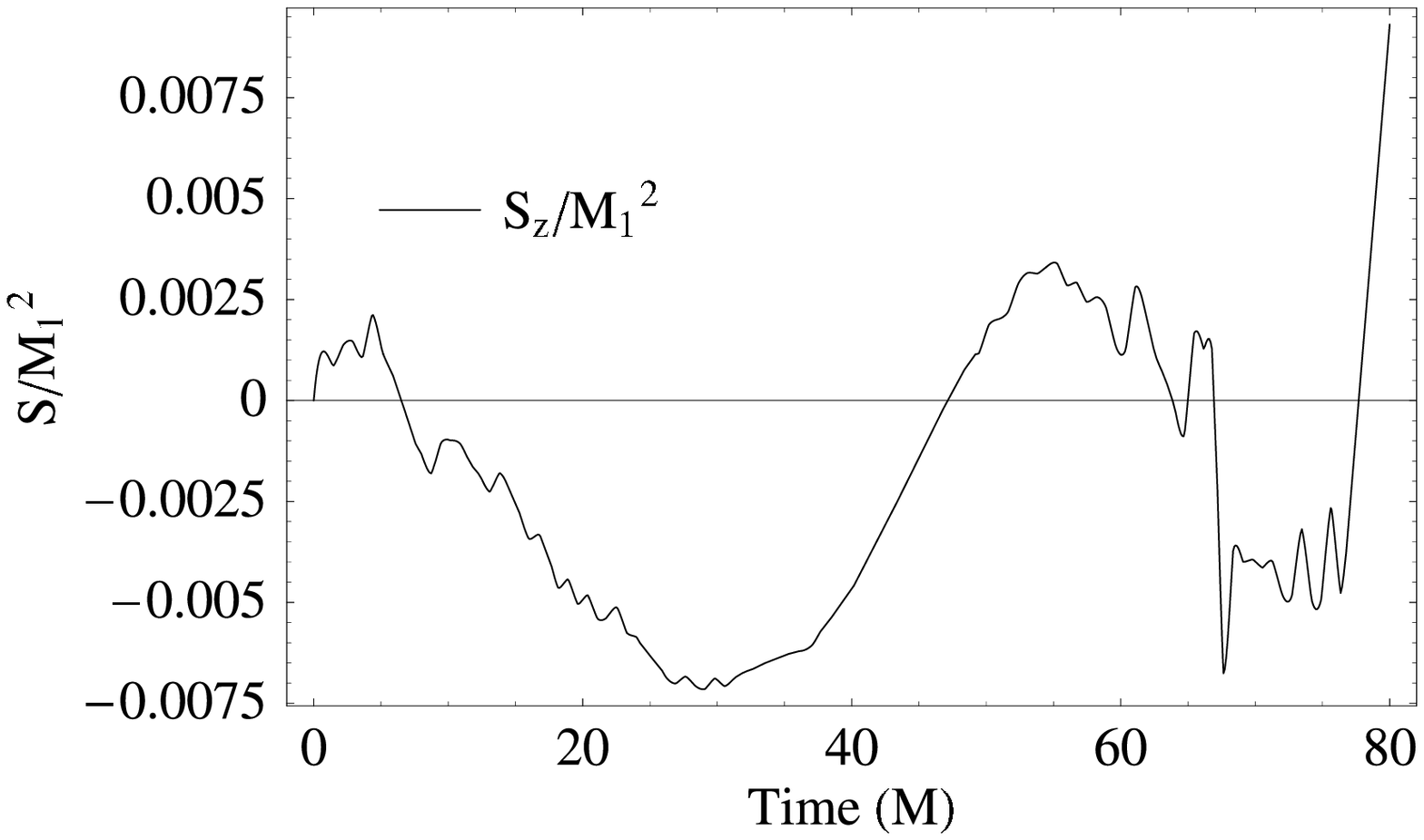}
\caption{\label{fig:alphaspins} 
Evolution of the spins $S_x$, $S_y$ and $S_z$ of one of the
black holes over the course of the
$\alpha=0$ simulation. The $x$- and $y$-components of the spin show noticeable
precession during the last orbit; the spin has rotated by $\pi/2$ after $80M$
of evolution. 
By contrast, the $z$-component (lower panel) displays small oscillations around
zero; these oscillations are not well resolved at the current accuracy of the
code. Merger occurs around (at $t\approx88M$),
at which time $S_x$ and $S_y$ drop to zero. The spin $S_z/M_f^2$ jumps
to a final value of $0.723$ corresponding to 
the merged black hole, reflecting the conversion of orbital angular
momentum into the spin of the final object. 
}
\end{figure}

The precession effects can be seen in Figure~\ref{fig:alphaspins}, which shows
$(S_x, S_y, S_z)$ as a function of time for the $\alpha = 0$ simulation
described earlier. Here the spins were calculated from the black holes' apparent
horizons, using the coordinate-based method outlined in
\cite{Campanelli:2006fy} and also used in \cite{Herrmann:2007ex}. 
We see that
the $x$- and $y$-components of the spin show noticeable precession during the
last orbit, and after $80M$ of evolution the spin vector has rotated by $\pi/2$. The
$z$-component shows small oscillations around zero, but these are not 
well resolved with the current accuracy of the code.
Note however by comparison with Fig.~\ref{fig:alphamotion} that the period of the oscillation
is roughly half an orbital period, which is consistent with the post-Newtonian
equations in Appendix~\ref{app:PNspins}. We will further comment on the comparison of the
spin precession with PN-results in Sec.~\ref{sec:PNcomparison}.
After the formation of  a common apparent horizon, at about 
$t\approx88M$, the $S_x$ and $S_y$ quickly drop to zero, and $S_z$ jumps
to its final value, corresponding to the spin of the final black hole, 
$S_z/M_f^2 \approx 0.7$, where $M_f$ is the mass of the final black hole.
We also compute the spin of the final black hole from quasi-normal ringdown
in Sec.~\ref{sec:finalspin} as $S_z/M_f^2 \approx 0.69$. 
Note that this value of the final spin is also the value for non-spinning equal-mass
inspirals, see e.g.~\cite{Bruegmann:2006at}.

\begin{figure}[ht]
\includegraphics[width=8cm]{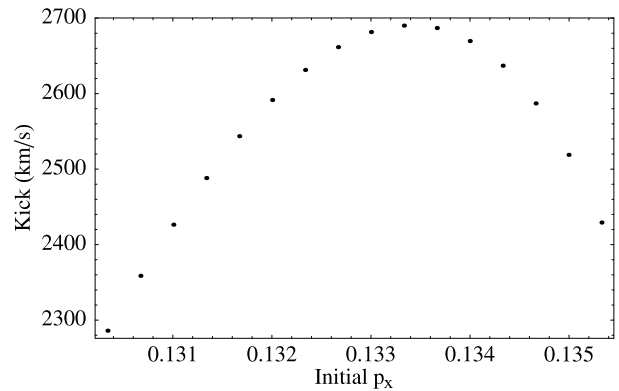}
\caption{\label{fig:pkicks} 
The final kick as a function of the initial momenta of the black holes for
simulations with the initial spin angle $\alpha = 0$. The differences between
the smallest and largest merger time is only about $15M$; these differences allow a
little more or less spin precession, and therefore have a strong effect on the
final recoil, which varies between $2300$ and $2700$ km/s.
}
\end{figure}

An example for the dependence of the final kick on parameters besides
$\alpha$, Figure~\ref{fig:pkicks} shows the final kick for $\alpha =
0$ simulations with differing values of the initial momenta of the
black holes (the P-series in Table~\ref{tab:parameters}). Changing the
initial momenta causes the merger time to change, and this means that
the spins have more or less time to evolve, and are therefore in
different directions when the black holes merge. The initial spin
directions are, however, the same for all of these simulations.

This leads us back to Eq.\ (1) in \cite{Campanelli:2007cg}, which was
originally written in terms of the spin angles at merger.
Above we formulated
Eq.~(\ref{eqn:trigkick}) in terms of the initial angle, but it would hold
equally well if, instead of choosing $\alpha = \alpha(t=0)$ we were to choose
some fixed $t_0$ and use $\alpha = \alpha(t = t_0)$ in
Eq.~(\ref{eqn:trigkick}). 
Only the phase constant $\alpha_0$ would change. 
The results shown in Figure~\ref{fig:pkicks}
suggest that we will also see oscillatory behavior if we make
a plot of the final kick versus merger time for a series of runs with the same
initial spin directions. 

In conclusion, given some superkick initial configuration we need to know
both the spin angle $\alpha$ and the time until merger in order to predict
the magnitude of the final kick from initial data.

\subsection{Duration of the recoil}
\label{sec:duration}

Simplified models aside, we know that the final kick is due
to an integration of $dP_z/dt$
over the entire evolution, and $dP_z/dt$ will be a complicated function of the
instantaneous spin directions. A post-Newtonian version of this function
is given in \cite{Kidder1995} and in Appendix~\ref{app:PNspins}.
Figure~\ref{fig:AlphadPzdt}
shows $dP_z/dt$ as a function of time for three simulations in the
$\alpha$-series. We make several observations about these plots. The main
contribution to the final kick originates from a time period of about  $60 M$.
Before that time the contribution is negligible, which is even clearer in
Figure~\ref{fig:R3motion}, which shows $dP_z/dt$ for the quasi-circular
model D6 which has a longer inspiral phase.
We also see in Figure~\ref{fig:AlphadPzdt} that
the function $dP_z/dt$ obtained for simulations with final
kick of $2500$, $0$ and $-2500~{\rm km/s}$ does not only
differ by a mere rescaling. Instead, the curve obtained for
$\alpha=\pi/2$ exhibits an additional oscillation.
\begin{figure}[ht]
\includegraphics[width=8cm]{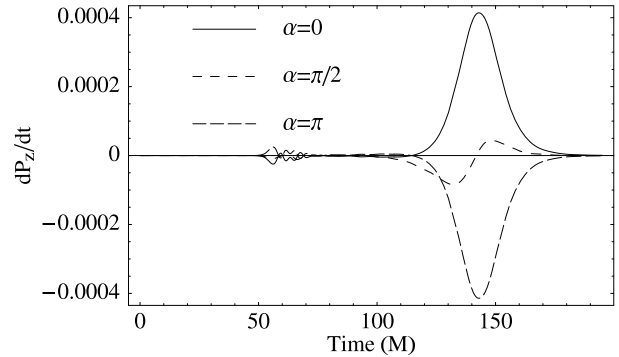}
\caption{\label{fig:AlphadPzdt} 
         Plot of $dP_z/dt$ as a function of time for the $\alpha = 0,\pi/2,\pi$
         simulations. Most of the linear momentum is radiated over a $60M$
         period of time, centered roughly around the merger time.
}
\end{figure}

\begin{figure}[ht]
\includegraphics[width=8cm]{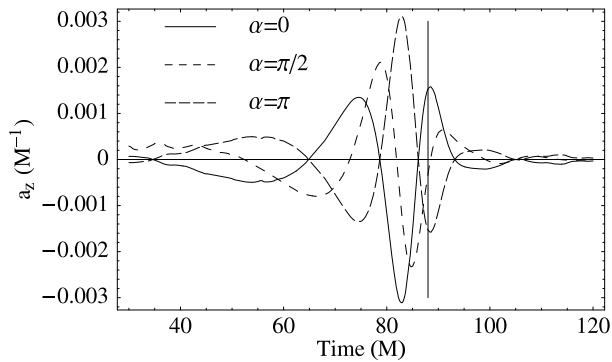}
\caption{\label{fig:AlphaAz} 
The coordinate acceleration of one of the punctures in the $z$-direction for the
$\alpha=0,\pi/2,\pi$ simulations. The vertical line represents the
time of first formation of a common apparent horizon. The curves
notably differ from those in Figure~\ref{fig:R3motion},
in particular they show significant extra oscillations,
while they have roughly the same
amplitude 
The plot essentially confirms
that the gauge dependence of the puncture
motion in this direction is not well understood (in contrast with the orbital
motion discussed in Appendix~\ref{app:PNmotion}), which prevents us from
drawing
quantitative conclusions about the kick velocity from the coordinate
acceleration.
}
\end{figure}

We would like to relate $dP_z/dt$ to the motion of the punctures, but
this is not trivial. The radiation is extracted at some radius $R_{ex}$, and
plotting this as a function of retarded time $r - R_{ex}$ gives only a crude
estimate of what is happening in the vicinity of the black holes at any given
time. One could try to improve this estimate by using instead the luminosity
distance (see for example \cite{Fiske05}), but we choose to simply look
at the puncture motion directly. We can calculate the coordinate acceleration of the
punctures in the $z$-direction, $a_z(t) = d^2 z(t)/dt^2$.
Figure~\ref{fig:AlphaAz} shows the acceleration of one of the punctures
in the $\alpha=0,\pi/2,\pi$ simulations. A vertical
line indicates the time at which a common apparent horizon forms, and thus
gives us an indication of how much of the motion is due to effects {\it before}
merger, and how much {\it after} merger.

Although most of the final kick is generated before merger, Figure~\ref{fig:AlphaAz}
suggests that $dP_z/dt$ is not negligible after the merger.
Referring back
to the plot of $dP_z/dt$ in Figure~\ref{fig:AlphadPzdt}, the waves from the
merger can be estimated to reach the radiation extraction sphere between
$138M$ (the time when the common apparent horizon forms, $88M$, plus
the extraction radius, $50M$) and $155M$ (when the final black hole's ringdown
can clearly be said to have begun, by observing that the wave amplitude has a
clean exponential decay).
Whatever time within this range we choose to denote
as ``merger'', it is clear that a significant contribution to the final
kick arises after that time.

\subsection{Comparison with post-Newtonian predictions}
\label{sec:PNcomparison}

In Figure~\ref{fig:AlphaAz} we see that there is a significant contribution to the kick after
the black holes merge. Before the merger, the main contribution comes from the
$30M$ before merger. During this time the black holes are at a separation of
$D < 3.5M$, and at this separation it is questionable whether an accurate
post-Newtonian description of the radiated linear momentum is meaningful. We
will now make a comparison between our numerical results and the predictions
from the 2.5-PN accurate expressions in \cite{Kidder1995}. 

Appendices \ref{app:PNspins}, \ref{app:PNmotion} and \ref{app:ADMTTtoHarmonic}
summarize the techniques we use to compare post-Newtonian and numerical
results. Briefly, Appendix \ref{app:PNspins} lists the expressions for the
spin evolution and radiated linear momenta as found in \cite{Kidder1995}. These
expressions were derived in the harmonic gauge. Our initial data are instead
in the ADMTT gauge (up to 2PN accuracy \cite{Jaranowski98a}), and although it
is not obvious how well we remain in the ADMTT gauge during evolution (but see \cite{Hannam:2007ik} for
a result that shows excellent agreement for larger separations), we
would like to see how much the results differ between the two
gauges. Appendix~\ref{app:ADMTTtoHarmonic} thus gives the expressions
necessary to transform the numerical quantities, which are assumed to be in
the ADMTT gauge, to the harmonic gauge. To do this we need to calculate the
momenta of the punctures as they evolve. Appendix~\ref{app:PNmotion} gives 2PN
expressions that relate the puncture's speeds (again assumed to be in the
ADMTT gauge) to momenta as given in Eq.~(\ref{eqn:vxPN}). We see in
Appendix~\ref{app:ADMTTtoHarmonic} that in fact the ADMTT $\rightarrow$ harmonic
transformation makes little difference to our results over the time when the 
PN approximation is valid. This result may not be surprising, but quantifies
any confusion that may arise when we compare results in the two gauges,
and eliminates any major concern that our results may change drastically
if we were to perform our simulations in the harmonic gauge.

As an aside, these formulas explain the speed at
which the punctures move in a black-hole binary moving-puncture simulation;
the puncture speeds and momenta are not related by the Newtonian formula $p =
mv$, but instead to good accuracy by its 2PN counterpart. 

\begin{figure}[ht]
\includegraphics[width=8cm]{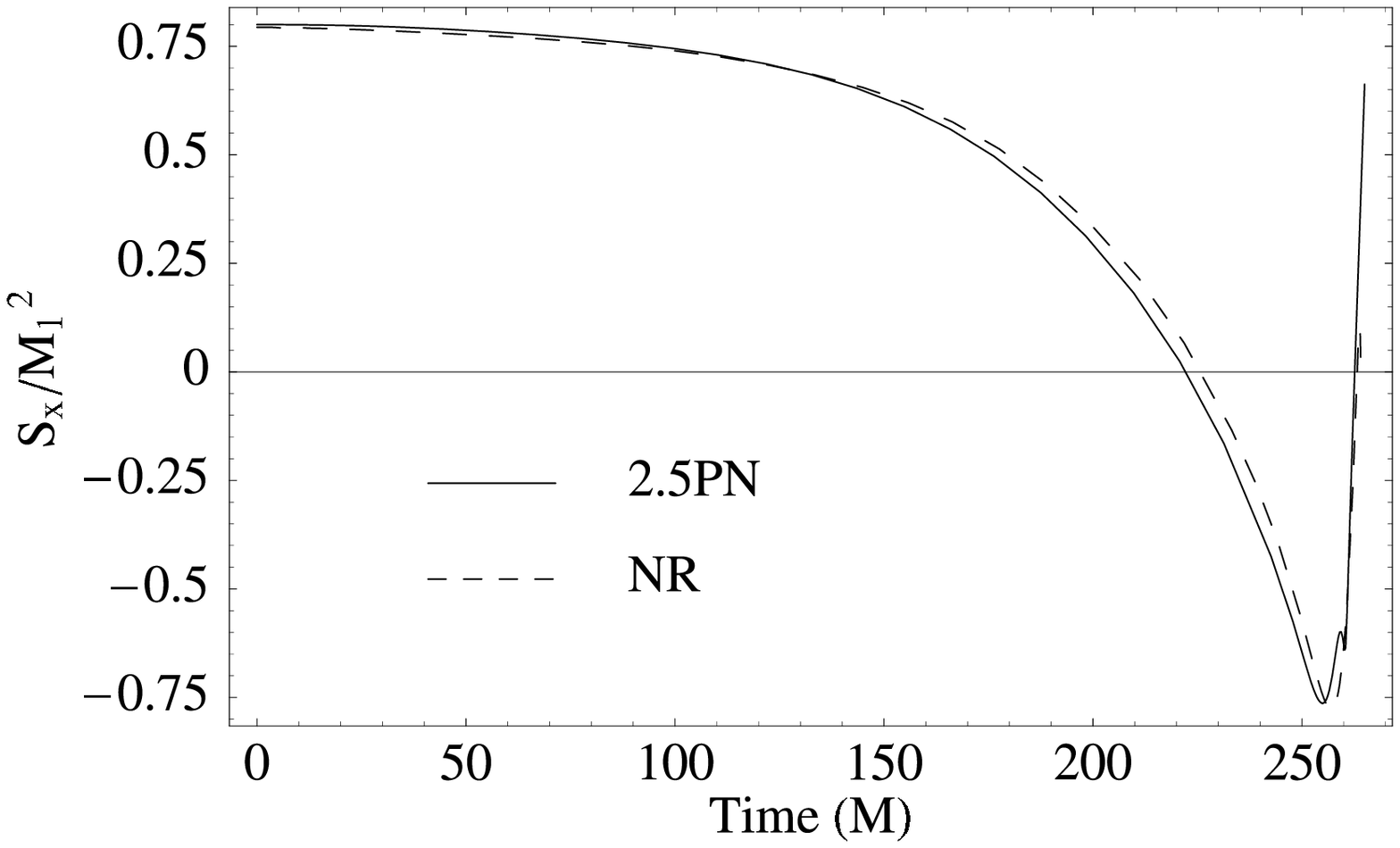}
\includegraphics[width=8cm]{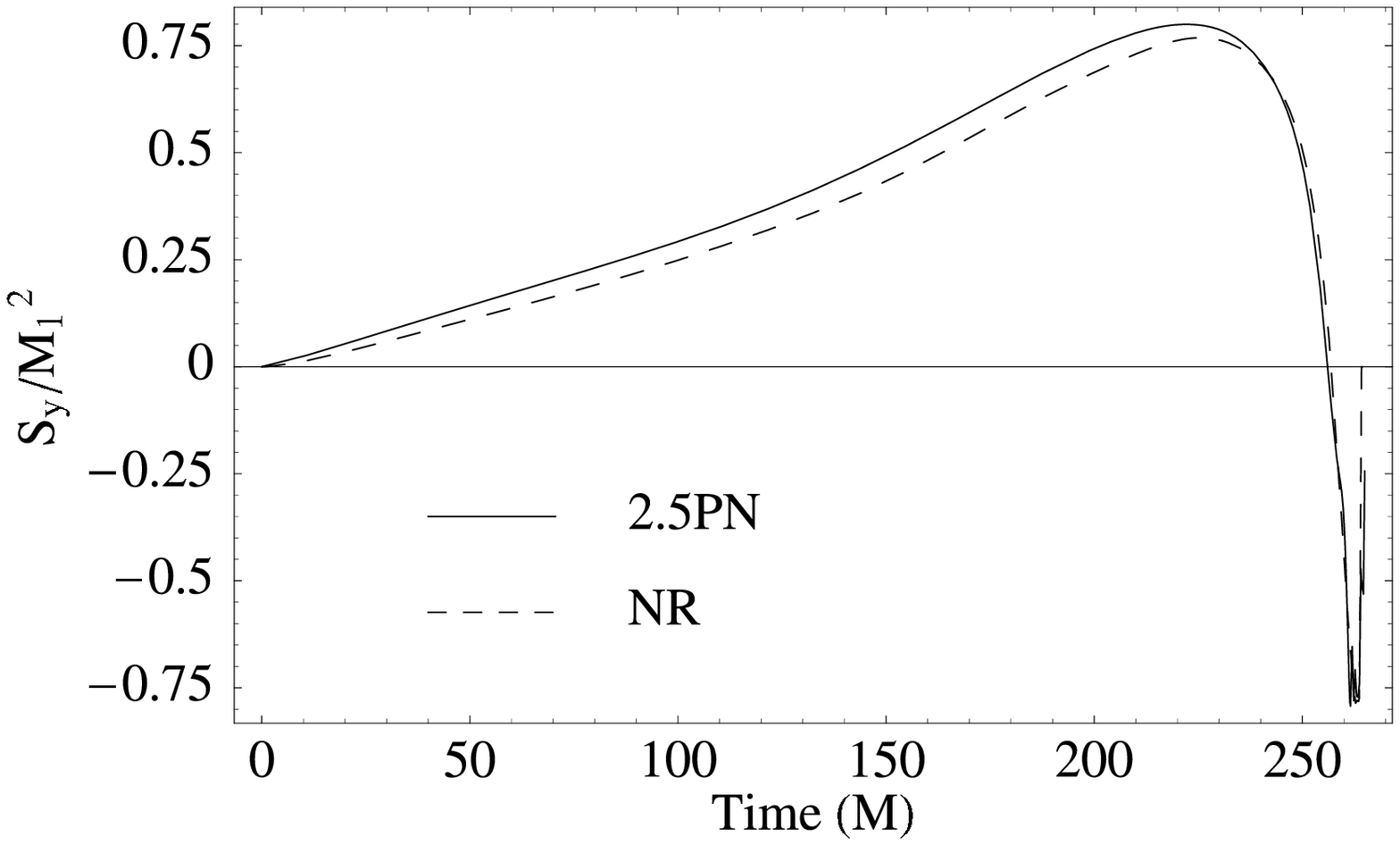}
\includegraphics[width=8cm]{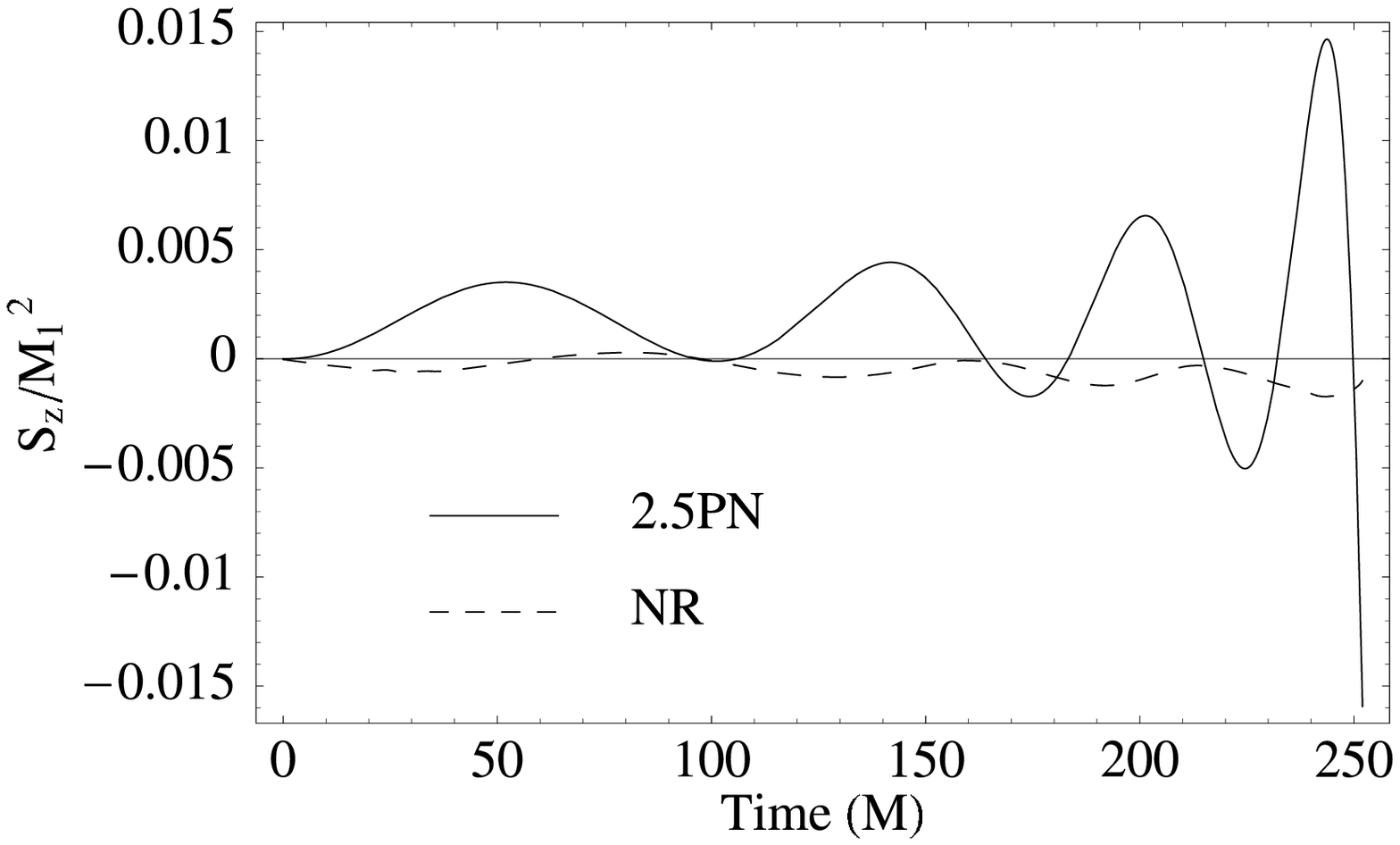}
\caption{\label{fig:KidderS} 
One black hole's spin as a function of time for simulation
D8. Also shown is the 2.5PN prediction for the spin evolution, with the
puncture dynamics $\{x_i,v_i\}$ used in the spin evolution equations
(\ref{eqn:KidderS}). The agreement is very good for $S_x$ and $S_y$, but poor
for $S_z$. This may once again be due to the numerical motion in the
$z$-direction being far more gauge-dependent than the motion in the $x$-$y$
plane. At late times $S_x$ and $S_y$ vanish, whereas $S_z$ is found to correspond to 
the angular momentum of the final black hole, compare Fig.~\ref{fig:alphaspins}
and Sec.~\ref{sec:finalspin}.
}
\end{figure}

Before considering the radiated linear momentum, we compare the post-Newtonian
predictions for the spin evolution with our numerical
results. Figure~\ref{fig:KidderS} shows the evolution of $S_x$, $S_y$ and
$S_z$ for simulation D8, compared with the 
predictions from Eqs.~(\ref{eqn:KidderS}). We see that there is
very good qualitative agreement in $S_x$ and $S_y$, even close to merger,
which occurs at around $t = 260M$. 
The $z$-component does not agree at all
well with the 2.5PN prediction, but we again note that the puncture motion in
the $z$-direction is much more gauge-dependent than the motion in the $x$-$y$
plane, and it is the positions and speeds of the punctures that we use when
evaluating the right-hand-sides of
Eqs.~(\ref{eqn:KidderS}). Furthermore, it is not clear how well the
numerical determination of the spin based on apparent horizons works in this context. Note
also that the {\em absolute} error is very small.
The frequency of the oscillations of the numerical simulation is rather
close to the PN result, which is approximately twice the orbital
frequency when precession effects are small (cmp.\ Appendix \ref{app:PNspins}). 
After merger, a few $M$ after the end of the figure, $S_z$ jumps to
its final value of around 0.7, and $S_x$ and $S_y$ drop to zero.

In the case of the radiated linear momentum flux $dP_z/dt$ and the final kick,
we know that the  assumptions underlying
the post-Newtonian expressions break down
when the black holes are very close. Eqs.~(\ref{eqn:KidderP}) diverge as
$1/r^5$ as the particles' separation $r \rightarrow 0$, so it is clear that a
sensible estimate of the kick cannot be made by simply integrating this
equation. What has been done in the past (see for example \cite{Favata2004})
is to assume a cut-off separation, and integrate the post-Newtonian
expression up to that point. We will now show that this approach is unlikely to give
correct results in the superkick case. 

\begin{figure}[ht]
\includegraphics[width=8cm]{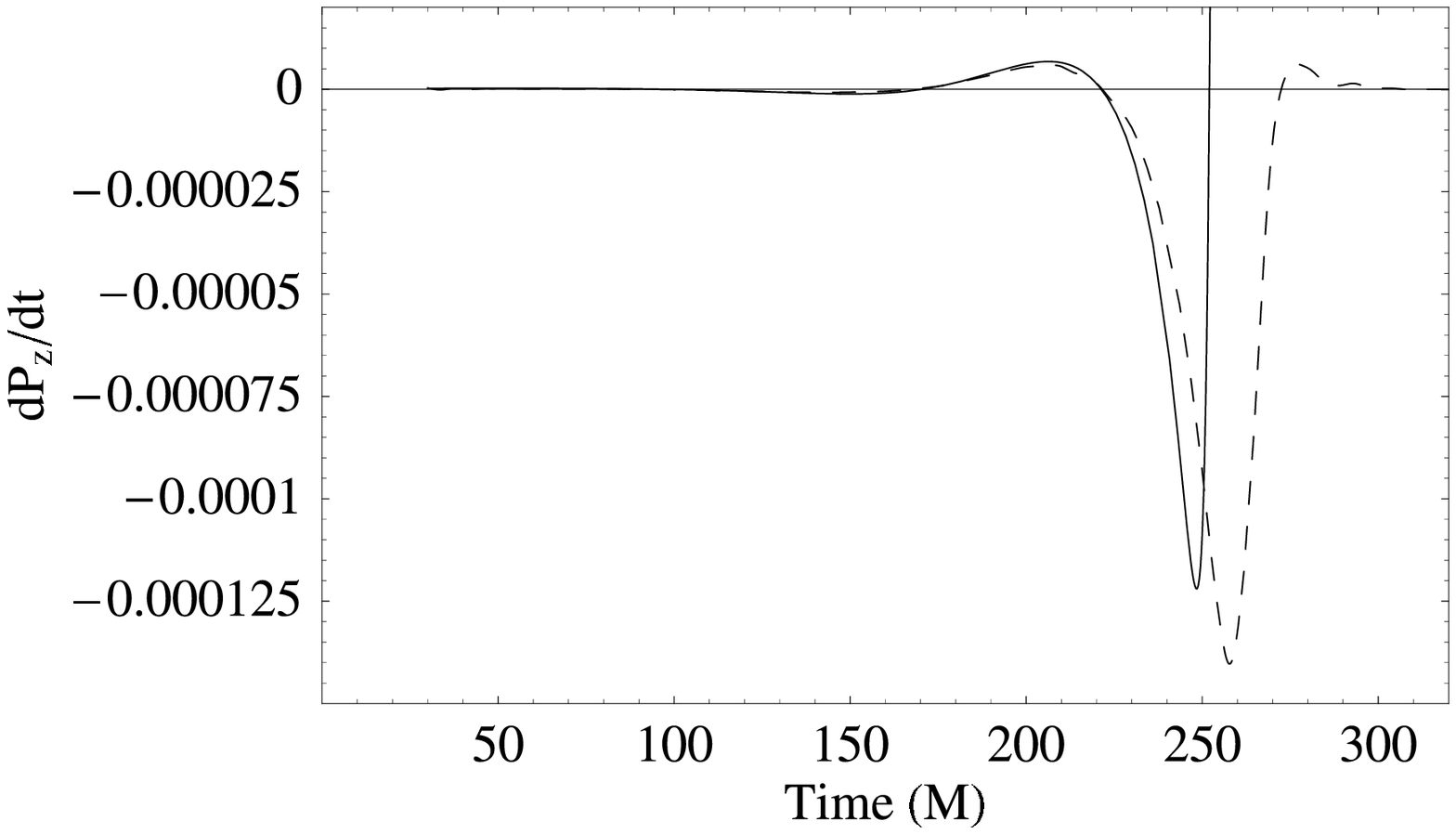}
\includegraphics[width=8cm]{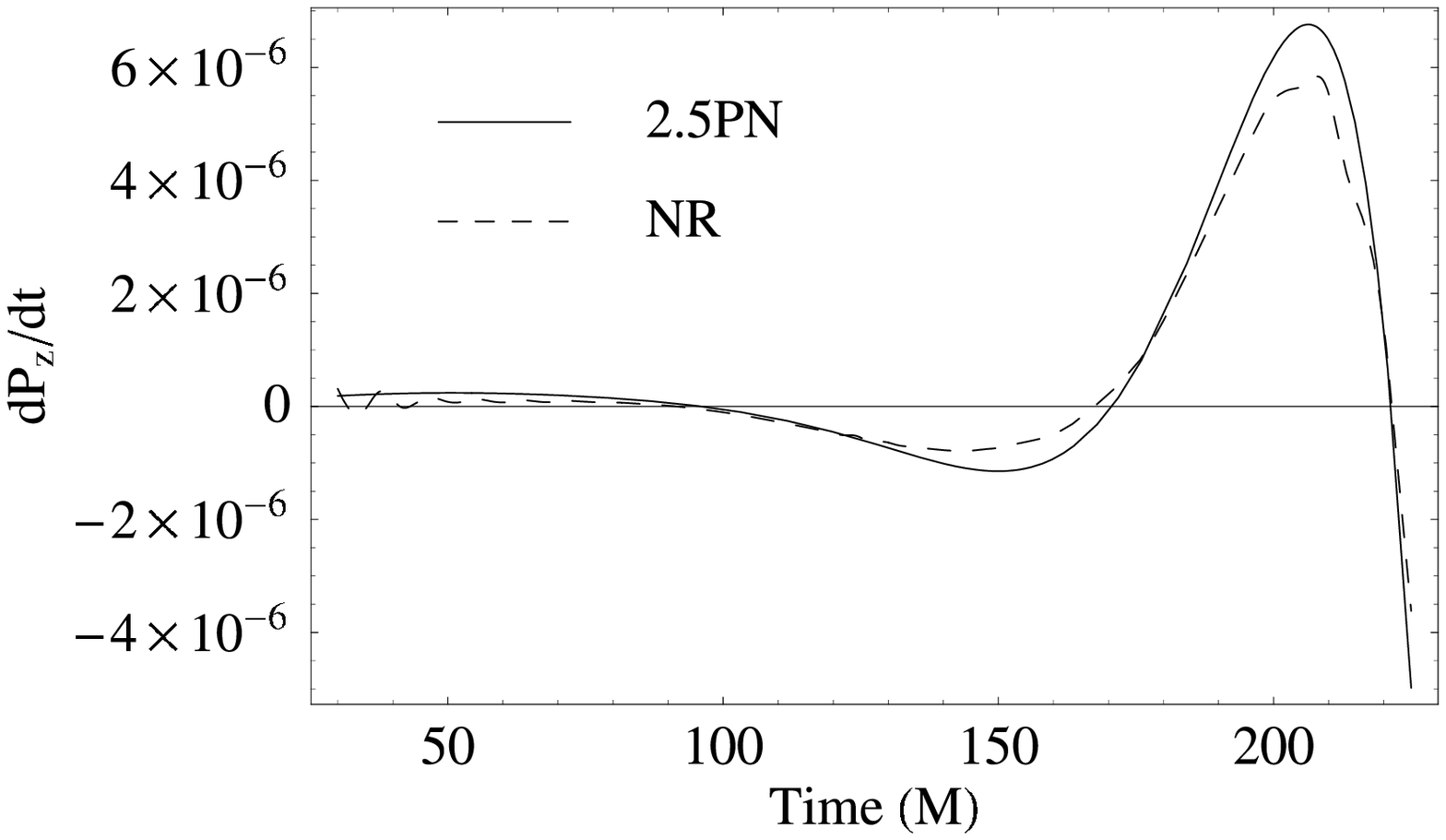}
\caption{\label{fig:KidderdPdt} 
Comparison of the numerical $dP_z/dt$ with that predicted by
Eqs.~(\ref{eqn:KidderP}). A time shift of $54.5M$ was applied to the
value calculated from the numerical wave extraction, to approximately take
into account the wave travel time between the punctures and the wave
extraction sphere by lining up the peak at $t \approx 200M$. The numerical relativity and 2.5PN 
results agree qualitatively at early times, but diverge quickly near merger, 
probably due to the $1/r^5$ term in Eq.~(\ref{eqn:KidderP}).  The lower plot
is a blow-up of the upper plot up to $t = 225M$. 
}
\end{figure}

Figure~\ref{fig:KidderdPdt} shows the function $dP_z/dt$ compared to the
numerical values for the D8 simulation at a retarded time $t - 54.5M$, chosen
to line up the early oscillations in $dP_z/dt$, and close to a naive guess of the
retarded time for the extraction radius $R_{ex} = 50M$. 
The post-Newtonian values of $dP_z/dt$ were calculated as
follows. Eqs.~(\ref{eqn:KidderP}) requires as input the positions, speeds
and spins of the two black 
holes. Rather than integrate the full post-Newtonian equations of motion, we
simply enter the appropriate quantities from a numerical evolution. This
allows us to compare, moment by moment, the post-Newtonian and numerical
predictions of $dP_z/dt$ for two particles (or black holes) with the $\{x_i,
p_i, S_i\}$ configuration. 

In Figure~\ref{fig:KidderdPdt} we once again see good
qualitative agreement at early times. At late times the post-Newtonian
prediction diverges, due to the $1/r^5$ term in
Eqs.~(\ref{eqn:KidderP}). What is most striking about this plot is that
the disagreement between numerical and post-Newtonian results becomes serious
around $50M$ before merger, which is just before the time when the major
contribution to the 
recoil begins in Figures~\ref{fig:R3motion} and \ref{fig:AlphaAz}. This
suggests that, at least in the special case of superkick configurations, if we
integrate the post-Newtonian $dP_z/dt$ up to the point where its accuracy
breaks down, we will grossly underestimate the value of the final kick. 

In order to accurately estimate the value of the kick analytically, one would
need to make a much more sophisticated choice of cutoff separation, and then
perhaps match to a close-limit analysis. Such a procedure was applied in
\cite{Sopuerta2006a,Sopuerta:2006et} to nonspinning binaries, and may well be
applicable in the spinning case. One may also be able to get good results from
a more careful post-Newtonian analysis, as was performed (also in the
nonspinning case) in \cite{Blanchet2005,Damour-Gopakumar-2006}. We would
expect that the superkick case would be an extreme and particularly
interesting test of such methods.

\subsection{Spin of the final black hole from ringdown}\label{sec:finalspin}

In Sec.~\ref{sec:precession} we found that the spin of the final black hole
as read off from the black hole horizon is $J_z/M^2 \approx 0.7$. Here we will also
determine the dimensionless Kerr spin parameter $a = J_z/M^2$ from the quasinormal ringdown
gravitational wave signal of the slowest decaying spin weighted 
{\em spheroidal harmonic} mode \cite{Detweiler1980,Leaver86}, which we measure by projecting it onto the 
$l=2, m=\pm 2$ spin weighted {\em spherical harmonics}.
These projected $l=2,m=\pm 2$ waveforms are split into amplitude and phase according to
$\psi_4 = A(t) \exp(i \varphi(t))$, 
we then perform analytical fits to the waveform for $170 \leq t/M \leq 230$,
where we see both a clean exponential decrease of the wave amplitude and a linear increase of 
the gravitational wave phase (corresponding to a constant frequency). 
Performing independent fits with a linear function for the wave phase and an exponential
for the amplitude we obtain values for the complex ringdown frequency $\omega_{QNM}$.
In order to factor out the overall mass scale we then perform a lookup of the dimensionless
quantity
$\mbox{Im}(\omega_{QNM})/\mbox{Re}(\omega_{QNM})$ (i.e. essentially the inverse quality factor)
in a table of QNM frequencies \cite{Berti06b}.

\begin{figure}[ht]
\includegraphics[width=9cm]{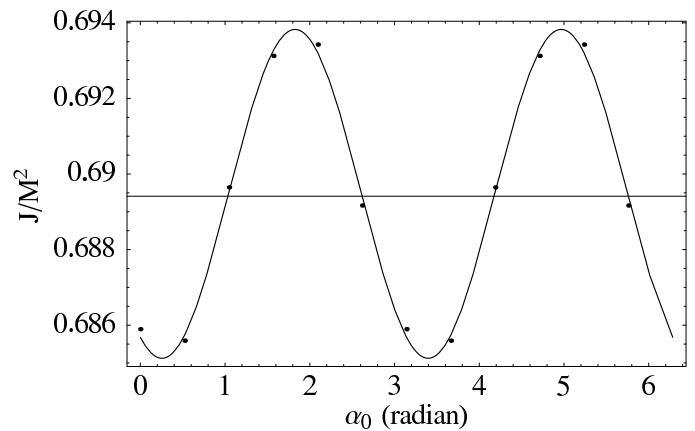}
\caption{\label{fig:aKerr}
The plot shows numerical results for $J/M^2 = \frac{a_{22} + a_{2-2}}{2}$ (points), obtained for extraction
radius $R_{ex} = 50 M$ for the final
black hole and an analytical fit (solid curve) -- in the text we conclude that
the final Kerr spin parameter does not show significant
variations (which might be further reduced by increased accuracy of the wave extraction).}
\end{figure}

We will consider in particular
data from the $\alpha$-series, see table (\ref{tab:parameters}). 
For the extraction radii
$R_{ex} = 30 M$, $50 M$, $75 M$ both the  $l=2,m=\pm 2$ results
can very well be fit with 
an analytic expression of the form $a_0 + a_1 \cos(\alpha + \varphi_1) + a_2
\cos(\alpha + \varphi_2)$, see Fig.~(\ref{fig:aKerr}). 
At each extraction radius we get consistent results for the amplitudes $a_0$,
$a_2$ and the phase shifts 
$\varphi_1$, $\varphi_2$ for the $m=-2,2$ modes, but we get the {\em opposite}
sign for $a_1$ for the $m=-2$ and $m=2$ modes. 
For $a_0$ we get  $(0.6963, 0.6891, 0.6891) \pm 5\times 10^{-4}$ (statistical error) for extraction radii $R_{ex} = (30 M, 50 M, 75 M)$.
For the oscillation amplitudes we get consistent results of $a_1 = 0.004 \pm 0.001$, $a_2 = - 0.004 \pm 0.001$, with
statistical errors corresponding to the 95 \% confidence interval and rounded to one significant digit.
We conclude that the asymptotic value of the dimensionless Kerr parameter is $a/M \approx 0.69$, 
which is consistent with the value $0.7$ we obtained from  the black hole horizon.
Since the oscillation $a_1 \cos(\alpha + \varphi_1)$, which has the periodicity of the kick velocity, is not consistent
between the $m=-2,2$ modes, and the oscillation $a_2 \cos(\alpha + \varphi_2)$ is of the same size, we conclude that
both may be non-physical, e.g. they could be due to gauge effects in the radiation extraction algorithm at
finite radius (we suggest an alternative explanation in the next paragraph). 
It is plausible that such problems are more serious in the present case of large kicks, than
when the black hole system does not move with respect to the center of gravity. It would be interesting
to analyze the present case more carefully e.g. along the lines discussed in \cite{Lehner:2007ip}.

Since the final spin of the black hole is close to the value for
non-spinning black hole mergers, it appears that the individual,
anti-aligned spins of the black holes do not contribute to the final
angular momentum, but rather cancel approximately during merger. This
is worth noting since based on the PN analysis and the numerical
evolutions there is a small oscillating $z$-component of the spin of
the individual black holes. At merger time, the $z$-component of the
black hole spins is added to the spin of the merged black holes due to
orbital motion. In principle it could happen that the initially small
$S_z$ is enlarged greatly (as the PN calculation becomes inaccurate),
e.g.\ it could happen that the separate spins precess significantly towards the
$z$-axis and add significantly to the final
angular momentum of the black hole. But this does not seem to happen, at best
there is a small positive or negative contribution to the final spin
depending on the momentary phase of the $S_z$ oscillation during
merger, e.g. as we see in Fig.~\ref{fig:aKerr}.

\section{Effects of recoil on SNR and template match}\label{sec:FF}

The gravitational recoil is essentially due to the symmetry-breaking 
between the dominating modes $l=2,m=\pm 2$. A natural question is how this symmetry breaking
is reflected in the overlap integrals of the gravitational waveforms. 
If the symmetry was not broken, the gravitational wave signal emitted towards  the ``north pole''
would provide the best template also for the south pole. Similarly, for a sequence of
waveforms that correspond to initial data that only differ in spin orientation (which
we have parameterized by the angle $\alpha$), we can ask how much signal-to-noise ratio (SNR)
is lost when trying to detect the gravitational wave signal corresponding to some value
of $\alpha$ with a template corresponding to a different value of $\alpha$.

Answering this question requires accurate waveforms, since any mismatch of waveforms
can also be due to lack of numerical resolution, errors from the finite extraction radius (which
can be significant, in particular because the recoil velocity creates an asymmetry of
the geometry of the extraction sphere), and the contribution of the initial junk radiation. 

For the data we are considering in this paper, the  initial junk radiation cannot be separated
from the main signal in a sufficiently clean fashion, neither can we obtain
accurate error bars on the wave signals of the whole $\alpha$-series to really settle the 
above questions.
Nevertheless some preliminary results will illustrate the issue.

We define the correlation function between two time series $x(t)$ and $y(t)$ for
a time shift $\tau$ as:
\begin{equation}
  \label{eq:correlation_function}
  R_{xy}(\tau) = \int_{-\infty}^{\infty}x(t) y^\star(t-\tau)dt\,,
\end{equation}
where a $^\star$ denotes complex conjugation.
Working with Fourier transforms
\begin{equation}
  \label{eq:FT}
  x(t) = \int_{-\infty}^\infty \tilde{x}(f)e^{2\pi ift}df\,.
\end{equation}
the correlation function can be written as 
\begin{equation}
  \label{eq:correlation_function_Fourier}
R_{xy}(\tau)  = \int_{-\infty}^{\infty}\tilde{x}(f) \tilde{y}^\star(f) e^{2\pi
    if\tau}df\,
\end{equation}
in terms of Fourier transforms $\tilde{x}(f), \tilde{y}(f)$ of the time series.
The function $R_{xy}(\tau)$ is thus simply the inverse Fourier transform of
$\tilde{x}(f)\tilde{y}^\star(f)$. The value of $\tau$ for which $R_{xy}$ is
maximal determines the time shift required to get the maximum
correlation between $x(t)$ and $y(t)$. 
The self-correlation $R_{xx}$ is maximal for $\tau=0$:
\begin{equation}
  \label{eq:self_correlation}
  R_{xx}(0) = (x|x) := \int_{-\infty}^\infty |\tilde{x}(f)|^2 df \,.
\end{equation}
More generally we can define the scalar product
\begin{equation}
  \label{eq:scalar_prod}
  (x|y) := \int_{-\infty}^\infty  \tilde x(f) \tilde y^\star (f)  df\,.
\end{equation}
The ``match'' which determines the efficiency of a template
$y$ to identify a signal $x$ is defined as
\begin{equation}
M = \max_\tau \frac{\vert R_{xy}(\tau) \vert}{\sqrt{ (x|x) (y|y) }}.
\end{equation}
Note that in gravitational wave data analysis
the orientation of a single detector actually reduces the signal to a real time series.
We can now evaluate $M$ for signals corresponding to different values of the initial spin-angle
$\alpha$ in our $\alpha$-series, or for signals corresponding to different angles in the sky
for a given value of $\alpha$, say one with a large value of the recoil. From symmetry we expect that
the mismatch $1-M$ when comparing the signals corresponding to the maximal
difference in the kick within the  $\alpha$-series ($\approx$ 5000 km/s) equals the mismatch
for the signals that correspond to the north and south poles for the maximal recoil case.
Indeed we find a value of $M \approx 0.94 \pm 0.01$ both for the $\alpha=0$ case, which is close
to the maximal recoil and for the maximal mismatch case within the  $\alpha$-series. 
Deviations of $\pm 0.01$ here come from comparing either the full complex waveform, or 
just $h_+$ or  $h_\times$. 
The uncertainties due to initial junk radiation, finite extraction radius and numerical error
may however be larger than 1 \%.
More accurate data than presented here will be required for conclusive error estimates.
Also, a detailed discussion of the
dependence of the radiation signal on the angle $\alpha$ and the consequences 
for gravitational wave data analysis is beyond the scope of the present paper.

\begin{figure}[ht]
\includegraphics[width=9cm]{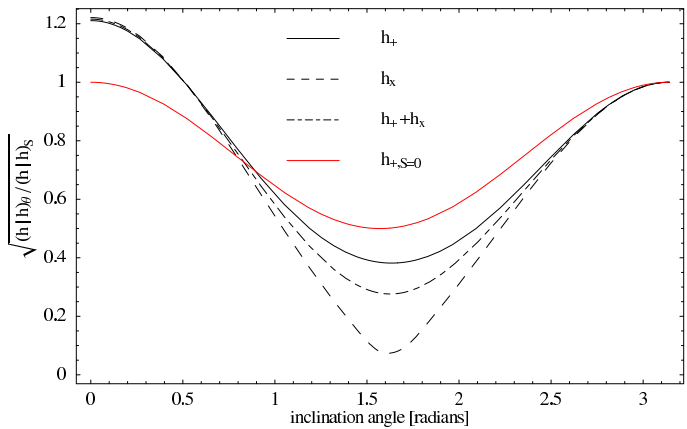}
\caption{\label{fig:brightness}
The dependence of the expression (\ref{eq:brightness}) is plotted as a function
of the inclination angle $\theta$ for  $h_+$, $h_\times$ and  $h_+ - i h_\times$ for the
near extremal member of the $\alpha$-series $\alpha=0$. For comparison we also show
the curve for the case without spin, when $h_+$ is symmetric around $\theta=\pi/2$.
The excess of signal toward the north pole compared with the south pole is roughly 25 \%.
}
\end{figure}

A related question is how much brighter the source appears in the direction opposite to
the recoil -- in which more radiation is emitted.
For the case of white noise we estimate the relative increase
in SNR computing the ratios of the norm of the strain $h$ for a given inclination
angle $\theta $ to the strain measured at the south pole ($\theta=\pi$) by computing:
\begin{equation}\label{eq:brightness}
\frac{\mbox{SNR}(\theta)}{\mbox{SNR}(\theta=\pi)} = 
\sqrt {\frac{(h(\theta) | h(\theta)}{(h(\theta=\pi) | h(\theta=\pi)} }\,.
\end{equation}
In Fig.~\ref{fig:brightness} we plot this ratio for the close-to-extreme case member
of the $\alpha$-series $\alpha=0$ for $h_+$, $h_\times$ and  $h_+ - i h_\times$.
The excess of signal toward the north pole compared with the south pole is roughly 25 \%.

\section{Discussion}\label{sec:discuss}

We have discussed ``superkick'' configurations, i.e., two equal-mass black holes with spins
anti-aligned and in the orbital plane,
as a simple but extreme ``test case'' for phenomena associated with the large recoil velocities
produced by spinning black-hole binary systems.
The high degree of symmetry results in the gravitational wave signal being dominated
by the $l=2,m=\pm 2$ spherical harmonics, i.e., the recoil is with good accuracy
proportional to the difference of energies radiated into the $l=2,m=\pm 2$ modes, see 
Figure~(\ref{fig:kick_vs_DeltaE}).

The asymmetry here is rather strong and in the extreme case $E_{22}/E_{2-2} \approx 2.4$.
For gravitational wave detection the ratio of the amplitude of the strain $h$ is more interesting,
in the direction opposite to the recoil
we find an excess of roughly 25 \% larger amplitude in the maximum recoil case.

For the large kicks one observes in the ``superkick'' configuration, one should certainly
worry about the accuracy of the wave extraction. For the present paper
we have been interested in a qualitative discussion rather than very high accuracy, but we
point out that a procedure to improve the accuracy of wave extraction via the Newman-Penrose
scalar $\Psi_4$ at finite radius has been discussed recently in \cite{Lehner:2007ip}. An
overall improvement in the accuracy of spinning black-hole binary simulations should also
be possible by employing higher-order spatial finite differencing \cite{Husa2007a} and using initial
parameters based on PN inspiral calculations \cite{Husa:2007ec}.

The main emphasis of this paper has been the comparison of the dynamics with post-Newtonian
predictions. We have found that the 2.5PN-accurate expressions given in \cite{Kidder1995}
accurately describe the spin evolution and linear momentum radiation up to about
$60M$ before merger. After that time the PN estimate of $dP_z/dt$ diverges from the 
numerical values. It is also after that time that we find the main contribution to the final
kick of the merged black hole, and this explains why it is difficult to make accurate predictions
of the kick by integrating the PN equations up to a cutoff separation.
In order to accurately analytically model the recoil for
superkick configurations (and possibly spinning black-hole binary configurations in 
general) we suggest that a more sophisticated post-Newtonian treatment would be
necessary, or a matching of PN methods during the early inspiral with a close-limit
analysis of the merger and ringdown. It has recently been found \cite{Herrmann:2007ex} 
that a phenomenological formula for the final kick, based on the angular dependence
of the terms in Eq.~(\ref{eqn:KidderP}), matches numerical data reasonably well. Having
found that the precise form of (\ref{eqn:KidderP}) fails to predict the linear momentum radiation
in the regime when the majority of the linear momentum is radiated, it will be interesting
to see how well such a phenomenological formula works for more general configurations,
or if a more detailed analytic study will suggest a more generally applicable formula.

\acknowledgments
We are grateful to P.~Ajith, B.~Krishnan and A.~Sintes for discussions regarding
material in Sec.\ \ref{sec:FF}, and to G.~Sch\"afer for sharing insights on the post-Newtonian approach. 
This work was supported in part by DFG grant SFB/Transregio~7
``Gravitational Wave Astronomy''.  We thank the DEISA Consortium
(co-funded by the EU, FP6 project 508830), for support within the
DEISA Extreme Computing Initiative (www.deisa.org).  Computations were
performed at LRZ Munich and the Doppler and Kepler clusters at the
Institute of Theoretical Physics of the University of Jena.

\begin{appendix}

\section{Initial-data parameters}
\label{app:qc}

In the Bowen-York/puncture data that we use, we must choose parameters for the
masses, separations, spins and linear momenta of the two punctures. Most of
the simulations studied in this paper are based on the MI configuration in
\cite{Gonzalez:2007hi}, for which the momenta were chosen without any
attempt to have the punctures move in quasi-circular orbit (although the
eccentricity in the resulting evolutions appears to be small). Often, however,
one wishes to produce quasicircular orbits, and the momenta for our 
non-MI-based simulations are chosen to meet that requirement. In this Appendix
we describe our procedure for calculating those parameters.
Note that in \cite{Husa:2007ec} we have described a procedure to further improve
the ``circularity'' of the initial data by using parameters obtained
from post-Newtonian inspirals.

In \cite{Bruegmann:2006at} we showed that a 3PN-accurate formula is sufficient
to calculate initial momenta for nonspinning binaries. In the spinning case we
make use of the results of Kidder \cite{Kidder1995}. They are in harmonic
coordinates, but as we will see in Appendix \ref{app:ADMTTtoHarmonic}, the difference
between harmonic coordinates and the ADMTT gauge that we expect our evolutions
to be in are small. Kidder's Eq.~(4.7) gives the
orbital angular momentum of a binary in circular orbit as 
\begin{widetext}
\begin{eqnarray}
\mathbf{L} & = & \mu (M r)^{1/2} \hat{\mathbf{L}}_N \left\{ 1 + 2 \left(
      \frac{M}{r} \right) - \frac{1}{4} \sum_{i=1,2} \left[ \chi_i
      (\hat{\mathbf{L}}_N \cdot \hat{\mathbf{S}}_i ) \left( 8
        \frac{M_i^2}{M^2} + 7 \eta \right) \right] \left( \frac{M}{r}
    \right)^{3/2} \right. \nonumber \\ 
& & \left.
+ \left[ \frac{1}{2} (5 - 9 \eta) - \frac{3}{4} \eta \chi_1 \chi_2
    \left[ (\hat{\mathbf{S}}_1 \cdot \hat{\mathbf{S}}_2 ) - 3 (
      \hat{\mathbf{L}}_N \cdot \hat{\mathbf{S}}_1) ( \hat{\mathbf{L}}_N \cdot
      \hat{\mathbf{S}}_2) \right] \right] \left( \frac{M}{r} \right)^2
\right\} \nonumber \\
& & - \frac{1}{4} \mu (M r)^{1/2} \sum_{i=1,2} \left[ \chi_i
  \hat{\mathbf{S}}_i \left( 4 \frac{M_i^2}{M^2} + \eta \right) \right] \left(
  \frac{M}{r} \right)^{3/2}, \label{eqn:KidderJ}
\end{eqnarray}
\end{widetext} and the total angular momentum is $\mathbf{J} = \mathbf{L} +
\mathbf{S}$. The variables in Eq.~(\ref{eqn:KidderJ}) are as follows. The
black holes are separated by a distance $r$ and have masses $M_i$, 
the total mass is $M = M_1 + M_2$, and the mass
ratio quantities are $\mu = M_1 M_2 / M$ and $\eta = \mu / M$. The black holes
have spins $\mathbf{S}_i$, and $\chi_i = |S_i| / M_i^2$. The quantity
$\hat{\mathbf{L}}_N$ denotes the unit vector in the direction of the angular
momentum of a Newtonian system of nonspinning particles. It need not point in
the same direction as the full orbital angular momentum $\mathbf{L}$ of the
system, and we may exploit this freedom to uniquely find a momentum
$\mathbf{P}$ that satisfies
\begin{equation}
\mathbf{L} = \mu (\mathbf{r} \times \mathbf{P}).
\end{equation} 

The specific setup of our data is as follows. The punctures are placed on the
$y$ axis and given momenta in the $x$ direction. The orbital angular momentum
therefore has only one component and that points in the $z$ direction. In
cases where the last term in Eq.~(\ref{eqn:KidderJ}) has a component in the $x$
or $y$ directions, we tilt $\hat{\mathbf{L}}_N$ such that the last term is
canceled out and $\mathbf{L} = L \hat{\mathbf{z}}$. Such a case will not arise
in the situations considered in this paper; the last term in
Eq.~(\ref{eqn:KidderJ}) will always sum to zero and we can simply write $p_x
= \mp L / r$.

\section{post-Newtonian treatment of spinning binaries}
\label{app:PNspins}

Consider two particles with masses $M_1$ and $M_2$, spins $\mathbf{S}_1$ and
$\mathbf{S}_2$, located at positions $\mathbf{x}_1$ and $\mathbf{x}_2$. We
define $\mathbf{x} = \mathbf{x}_1 - \mathbf{x}_2$ and $\mathbf{v} = d
\mathbf{x}/dt$. Expressions for the evolution of the spins are given up to 2.5
post-Newtonian order in harmonic coordinates by Kidder \cite{Kidder1995},
\begin{eqnarray} 
\dot{\mathbf{S}}_1 & = & \frac{1}{r^3} \left\{ (\mathbf{L}_N \times
  \mathbf{S}_1) \left( 2 + \frac{3}{2} \frac{M_2}{M_1} \right) - \mathbf{S}_2 \times
  \mathbf{S}_1 \right. \nonumber \\
& & \left. + 3 ( \hat{\mathbf{n}} \cdot \mathbf{S}_2 ) \mathbf{n}
\times \mathbf{S}_1 \right\}, \nonumber \\
\dot{\mathbf{S}}_2 & = & \frac{1}{r^3} \left\{ (\mathbf{L}_N \times
  \mathbf{S}_2) \left( 2 + \frac{3}{2} \frac{M_1}{M_2} \right) - \mathbf{S}_1 \times
  \mathbf{S}_2 \right. \nonumber \\
& & \left. + 3 ( \hat{\mathbf{n}} \cdot \mathbf{S}_1 ) \mathbf{n}
\times \mathbf{S}_2 \right\},  \label{eqn:KidderS}
\end{eqnarray} where the Newtonian orbital angular momentum is given by
$\mathbf{L}_N = \mu ( \mathbf{x} \times \mathbf{v} )$. Note that a
particular spin supplementary condition has been chosen.

The radiated linear momentum is in turn given by Newtonian and spin-orbit
contributions, \begin{eqnarray}
\dot{\mathbf{P}}_N & = & - \frac{8}{105} \frac{\delta m}{M} \eta^2 \left(
  \frac{M}{r} \right)^4 \left\{ \dot{r} \hat{\mathbf{n}} \left[ 55 v^2 - 45
    \dot{r}^2 + 12 \frac{M}{r} \right] \right. \nonumber \\
&  & \left. + \mathbf{v} \left[ 38 \dot{r}^2 - 50 v^2 - 8 \frac{M}{r} \right]
\right\}, \\
\dot{\mathbf{P}}_{SO} & = & - \frac{8}{15} \frac{\mu^2 M}{r^5} \left\{ 4 \dot{r}
( \mathbf{v} \times \mathbf{\Delta} ) - 2 v^2 ( \hat{\mathbf{n}}
\times \mathbf{\Delta}
) \right. \nonumber \\
&  & \left. - (\hat{\mathbf{n}} \times \mathbf{v} ) [ 3 \dot{r} ( \hat{\mathbf{n}}
\cdot \mathbf{\Delta} ) + 2 (\mathbf{v} \cdot \mathbf{\Delta})] \right\},
\label{eqn:KidderP} 
\end{eqnarray} where $\mathbf{\Delta} = M (\mathbf{S}_2 / M_2 - \mathbf{S}_1 /
M_1 )$ and $\delta m = M_1 - M_2$. Clearly the Newtonian contribution, which
was used by Fitchett \cite{Fitchett1983} to provide an early estimate of the
recoil from the merger of nonspinning binaries, is zero in the equal-mass
case. 

To evaluate Eqs.~(\ref{eqn:KidderS}) and (\ref{eqn:KidderP}) one needs the
particles' positions and velocities as a function of time, i.e., one needs to
solve the post-Newtonian equations of motion. Alternatively, we can use the
motions of the punctures calculated in our numerical simulations. This allows
us to compare the precession of the spins and the radiated linear momentum
with that predicted by post-Newtonian theory for the same motion. The results
of this comparison are discussed in Section~\ref{sec:PNcomparison}.

It is instructive to reduce
Eqs.~(\ref{eqn:KidderS})--(\ref{eqn:KidderP}) to the special case of
equal mass and $\pi$-symmetry. Equal masses $M_1=M_2$ imply
$\delta m=0$ and hence $\dot{\mathbf{P}}_N = 0$. Furthermore, 
$\mu=M/4$, $\eta=1/4$, and 
$\mathbf{\Delta}=2 (\mathbf{S}_2 - \mathbf{S}_1)$. 

For $\pi$-symmetry (which implies equal masses), the positions of the punctures 
$\mathbf{x}_i=(x_i, y_i, z_i)$ satisfy
\begin{equation}
(x_2, y_2, z_2) = (-x_1,-y_1,+z_1)
\end{equation}
{\em for all times}. This implies for the relative position and velocity
variables
\begin{equation}
\mathbf{x} = (2x_1, 2y_1, 0), \quad \mathbf{v} = (2v_{1x},2v_{1y},0).
\end{equation}
In words, for $\pi$-symmetry the punctures can move in unison in the
$z$-direction, and in general $z_i(t)$ will describe an accelerated motion.
However, the PN equations (\ref{eqn:KidderS})--(\ref{eqn:KidderP}) are
expressed in terms of $\mathbf{x}$ and $\mathbf{v}$, which describe
the motion in coordinates in which the center of mass is at rest. 
Both $\mathbf{x}$ and $\mathbf{v}$ are orthogonal to the $z$-axis and
lie in the $z=0$ plane. In particular, $\mathbf{v}$ does not have a
component in the $z$-direction, so for general $z_i(t)$ the center of
mass frame is an accelerated frame. A related statement is that in
$\pi$-symmetry the orbital plane remains orthogonal to the $z$-axis,
there is no orbital plane precession, and 
$\mathbf{L}_N=\mu (\mathbf{x}\times\mathbf{v})$ is parallel to the
$z$-axis at all times.

For $\pi$-symmetry the spins can be written as 
\begin{equation}
\mathbf{S}_1=\mathbf{\zeta}+\mathbf{\sigma}, \quad 
\mathbf{S}_2=\mathbf{\zeta}-\mathbf{\sigma},
\end{equation}
where $\zeta$ and $\sigma$ are the components of the spin parallel and
orthogonal to the $z$-axis, respectively. Hence the sum of the spins
$\mathbf{S}=\mathbf{S}_1+\mathbf{S}_2$ points
in the $z$-direction and the weighted spin difference 
$\mathbf{\Delta}=2(\mathbf{S}_2 - \mathbf{S}_1)$ is orthogonal to the
$z$-axis, 
\begin{equation}
\mathbf{S}=2\mathbf{\zeta}, \quad \mathbf{\Delta} = -4\mathbf{\sigma}.
\end{equation}
For the initial data we choose $\mathbf{\zeta}=0$, but in general
$\mathbf{\zeta}$ is not constant in time.
The time derivative of the spins can be written as
\begin{equation}
\dot\mathbf{S}_i = \mathbf{\Omega}_j \times \mathbf{S}_i, \quad
\mathbf{\Omega}_j = \frac{1}{r^3} \left(\frac{7}{2} \mathbf{L}_N -
\mathbf{S}_j + 3 (\hat\mathbf{n} \cdot \mathbf{S}_j)\hat\mathbf{n}\right),
\end{equation}
where one part is precession about the $z$-axis due to
the orbit-spin term, $\mathbf{L}_N\times\mathbf{S}_i$, but the axis of
precession is in general not parallel to the $z$-axis due to the
spin-spin terms. For $\pi$-symmetry we obtain
\begin{eqnarray}
\dot\mathbf{S} &=& -\frac{6}{r^3} (\hat\mathbf{n} \cdot \mathbf{\sigma})
                                (\hat\mathbf{n} \times\mathbf{\sigma})
= -\frac{3}{r^3} |\sigma|^2 \sin(2\alpha) \, \hat\mathbf{z},
\end{eqnarray}
where $\hat\mathbf{z}$ is the unit vector in the $z$-direction and
$\alpha$ is the angle between $\hat\mathbf{n}$ and $\mathbf\sigma$.
The $z$-component of the spins, $\mathbf{\zeta}=\mathbf{S}/2$,
oscillates with $\sin(2\alpha(t))$. Since for negligible
precession $\alpha(t)$ is equal to the orbital phase plus a phase
shift, we expect two oscillations per orbit, which roughly 
agrees with observation, see Secs.~\ref{sec:precession} and \ref{sec:PNcomparison}. 
For the weighted difference of the spins
\begin{eqnarray}
\dot\mathbf{\Delta} = -4\dot\mathbf{\sigma} &=& 
- \frac{7M}{2r^2}(\hat\mathbf{n} \times\mathbf{v})\times
\mathbf{\sigma} 
\nonumber \\
&& -
\frac{4}{r^3} (2\mathbf{\sigma} + 
3 |\sigma| \cos(\alpha) \, \hat\mathbf{n}) \times \mathbf{\zeta},
\end{eqnarray}
which contains precession due to $\mathbf{L}_N$ at order $1/r^2$ and
the spin-spin term at order $1/r^3$. The term
$\mathbf{\zeta}\times\mathbf{\sigma}$ describes a modulation of the
precession about the $z$-axis since $\mathbf{\zeta}$ oscillates around
zero. 

For the radiated linear momentum we note that for
$\pi$-symmetry the three vectors $\hat\mathbf{n}$, $\mathbf{v}$, and
$\mathbf{\Delta}$ in (\ref{eqn:KidderP}) are orthogonal to the
$z$-axis, and hence $\dot\mathbf{P}_{SO}$ is parallel to the $z$-axis
as it should be. The angle between $\hat\mathbf{n}$ and $\mathbf{v}$
varies slowly over the entire inspiral from about $\pi/2$ for circular
orbits to a value less than $\pi$ for the plunge. The angle between
the orbital vectors and $\mathbf{\Delta}$ oscillates with the orbital
and precession time scales. Making the approximation that
$\hat\mathbf{n}\cdot\mathbf{v}\approx 0$ and
$\hat\mathbf{n}\times\mathbf{v}\approx v\hat\mathbf{z}$, we obtain
\begin{equation}
\dot\mathbf{P}_{SO} \approx - \frac{2}{15}\frac{M^3}{r^5}
\left(7 \dot r v (\hat\mathbf{n}\cdot\mathbf{\sigma}) +
 4 v^2 (\frac{\mathbf{v}}{v}\cdot\mathbf{\sigma})\right) \, \hat\mathbf{z}.
\end{equation}
Even for quasi-circular orbits where in addition we set $\dot r = 0$
there will be radiation of linear momentum in the $z$-direction, which
however averages to zero over time. 

Note that in general there are two
contributions, one proportional to $\dot r v$ and the other to $v^2$,
and they are offset in phase depending on the angles between the spin
and the orbital vectors. 
As the system approaches the plunge phase, the $\dot r v$ term should become
as important as the $v^2$ effects. Note that we have not discussed the
$\dot\mathbf{P}_{SS}$ spin-spin contribution at next PN order, 
which could also be examined for potentially large
contributions near the plunge, but in numerical simulations of head-on
collisions the resulting kicks have been found to be small \cite{Choi:2007ck}. 

The PN expressions (\ref{eqn:KidderP}) and the above discussion apply
in the regime where the post-Newtonian approximations are valid. We
see in Section~\ref{sec:PNcomparison} that these expressions describe 
the radiation of linear momentum with reasonable accuracy up to about
$50M$ before merger.

\section{PN calculation of puncture motion}
\label{app:PNmotion}

In moving-puncture simulations we can readily track the motion of the
punctures and record their positions $x(t)$ and velocities $v(t) = -
\beta(t)$. We may then be tempted to make a Newtonian analogy and guess that the
puncture's momentum is $P = M_1 v$ for a black hole with mass $M_1$. 
However, when we compare this to the
momentum specified in the initial data, the two values differ
significantly. For example, evolve two equal-mass punctures with initial
separation $D = 8M$, $P = 0.14$, $M_1 = M_2 = 0.5$. From the numerical
puncture motion we find $M_1 v \approx 0.075$; this value disagrees with $P$
by almost a factor of two. 

During a simulation the ``punctures'' are at an infinite proper distance from
their black holes' horizons \cite{Hannam:2006vv,Hannam:2006xw}, and we may
worry that correctly physically interpreting the punctures' motions requires a
thorough investigation of the gauge and geometry of the punctures as they evolve. In
fact, the punctures' motions can be understood from a simple post-Newtonian analysis. 

Up to 2PN order, the Hamiltonian for two point particles in the ADMTT gauge
and center-of-mass frame has been derived in \cite{Damour85,Damour88}.

From the Hamiltonian equations of motion, \begin{equation}
\dot{x}_i = \frac{\partial H}{\partial P_i},
\end{equation} where $x_i$ is the separation vector between the two
particles. At Newtonian order we recover $\dot{x}_i = P_i / (2 \mu)$. Up to
2PN order we have for circular orbits
\begin{eqnarray}
\dot{x}_i & = &  \frac{P_i}{2 \mu} \left\{ 1 - \frac{1}{c^2}\left( \frac{P^2 (1 -
    3\eta)}{2 \mu^2} + \frac{ M (3 + \eta) }{R}\right) \right. \nonumber \\
& & \left. + \frac{1}{c^4} \left( \frac{3 P^4 (1 - 5 \eta + 5 \eta^2)}{8 \mu^4} -
  \frac{M P^2 (-5 + 20 \eta + 3 \eta^2)}{ 2 \mu^2 R } \right. \right. \nonumber \\
& & \left. \left.\ \ \ \ \ \ \ \ + \frac{M^2 (5 + 8 \eta) }{ R^2 } \right)
\right\}  \label{eqn:vxPN} 
\end{eqnarray} The more general expressions (removing the assumption of
circular orbits), and the 3PN terms, will be omitted here for brevity. They
can be readily calculated from the Hamiltonian in \cite{Damour88}.

As an example, consider the $D = 8M$ quasi-circular orbit parameters from the
sequence presented in \cite{Tichy:2003qi}, for which $P = 0.111 M$ and the
orbital frequency is $M \Omega =  0.0376$. From the orbital frequency we can
calculate that the punctures will move at a speed $\dot{x} =  0.150$, which
is approximately equal to the observed value in a simulation. From
the momentum, the Newtonian prediction of the speed is $\dot{x}_N = 0.223$,
which is far too high. The 1PN prediction is $\dot{x}_{1PN} = 0.126$ (now
the value is too small), and the 2PN and 3PN predictions are $\dot{x}_{2PN} =
0.151$ and $\dot{x}_{3PN} = 0.149$. The 2PN and 3PN predictions are both
very close to the observed value.

In addition to providing a pleasing consistency between the dynamics observed
in moving-puncture simulations and that predicted by post-Newtonian theory,
this analysis is necessary when converting between ADMTT and harmonic
gauges in Appendix~\ref{app:ADMTTtoHarmonic}, where we will need to invert
equations like (\ref{eqn:ADMTTtoHarmonic}) to estimate the black holes' momenta as a
function of time from the puncture motion.

\section{ADMTT to Harmonic transformation}
\label{app:ADMTTtoHarmonic}

Our initial data are, up to 2PN order, in the ADMTT gauge
\cite{Jaranowski98a}. The post-Newtonian expressions for spin evolution and
linear momentum radiation listed in Appendix~\ref{app:PNspins} are in the
harmonic gauge. Although it is not clear how closely our evolved data adhere
to the ADMTT gauge, it is nonetheless useful to assume that they remain in the
ADMTT gauge and transform the results to the harmonic gauge and see how
different they are. 

A transformation between ADMTT and harmonic coordinates is provided up to 2PN
order by Damour and Sch\"{a}fer \cite{Damour85}. If $\mathbf{x}_i$ are the
ADMTT coordinates of the $i$-th particle and $\mathbf{X}_i$ are the
corresponding harmonic coordinates, then the transformation for a binary
system is \begin{eqnarray} 
\mathbf{X}_i & = & \mathbf{x}_i + M_i \left\{ \mathbf{n} \left( \frac{5}{8}
    v_j^2 - \frac{1}{8} (\mathbf{n} \cdot \mathbf{v}_j )^2 + \frac{7 M_i}{4R}
    + \frac{M_j}{4R} \right) \right. \nonumber \\
& & \left. + \left( \frac{1}{2} \mathbf{v}_i - \frac{7}{4} \mathbf{v}_j \right) (
\mathbf{n} \cdot \mathbf{v}_j) \right\}. \label{eqn:ADMTTtoHarmonic}
\end{eqnarray} The velocities $v_{\{a,b\}}$ in
Eq.~(\ref{eqn:ADMTTtoHarmonic}) are not the coordinate speeds, but are
instead $v_i = p_i / m_i$, and the momenta $p_i$ must be determined by the
procedure described in Appendix~\ref{app:PNmotion}. 

\begin{figure}[ht]
\includegraphics[width=8cm]{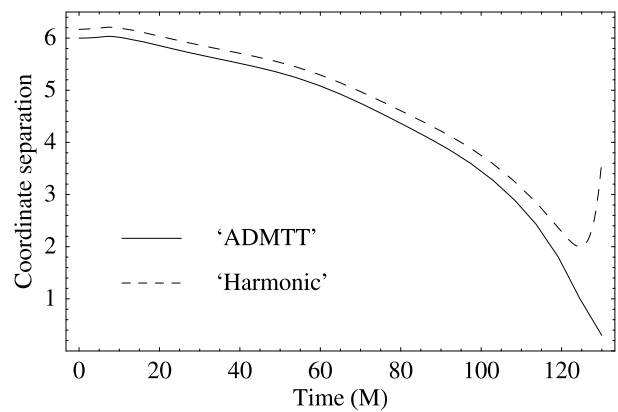}
\caption{\label{fig:Harmonic} 
Coordinate separation as a function of time for the D6 simulation, comparing
the numerical data (presumed to be in ADMTT coordinates), and the same
data transformed to harmonic coordinates. The difference is less than 10\%
up until about $15M$ before merger. It is also clear that after this time (when
the ``harmonic'' curve turns upward)
the PN approximations in the coordinate transformation break down. 
}
\end{figure}

\begin{figure}[ht]
\includegraphics[width=8cm]{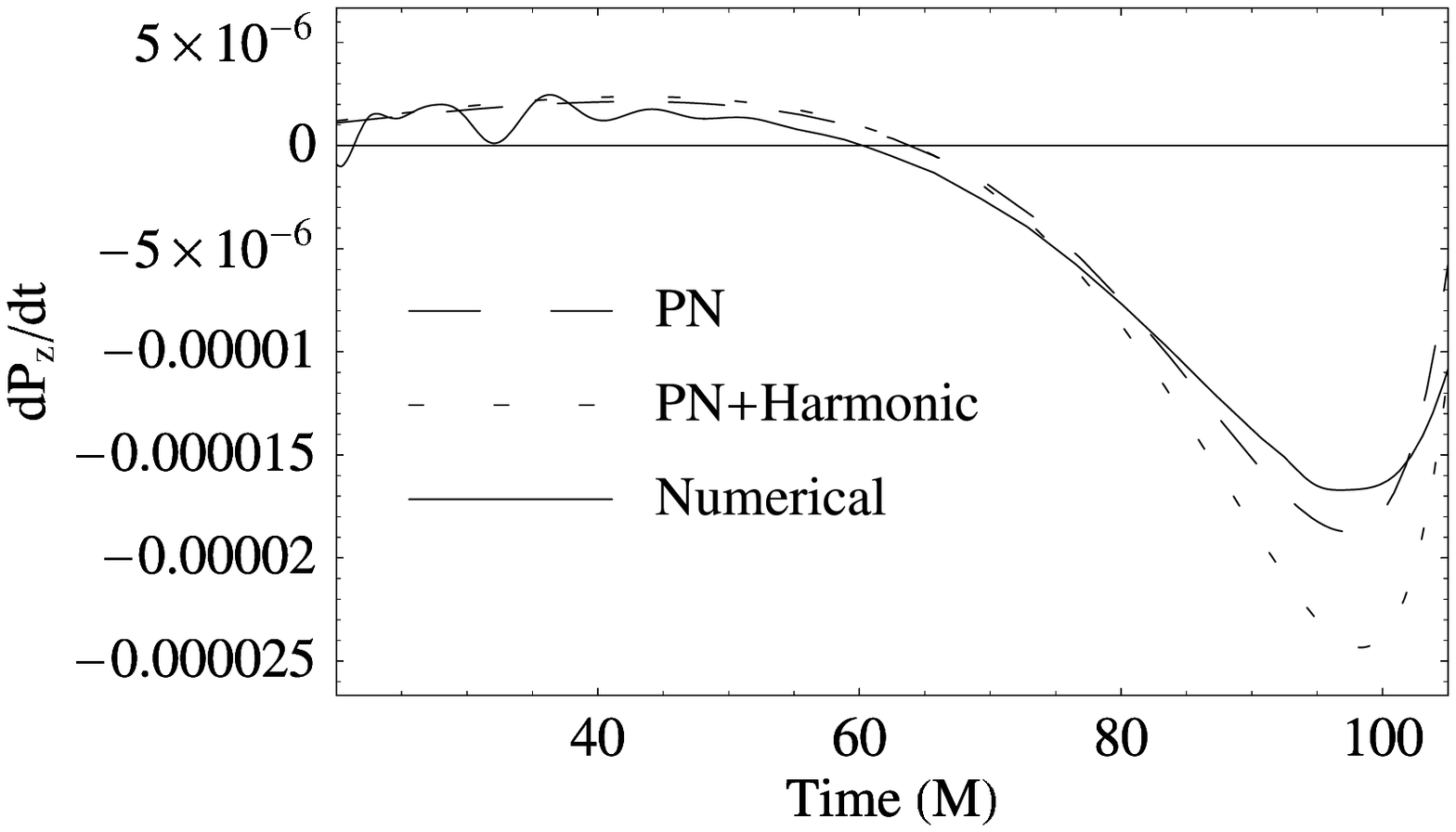}
\includegraphics[width=8cm]{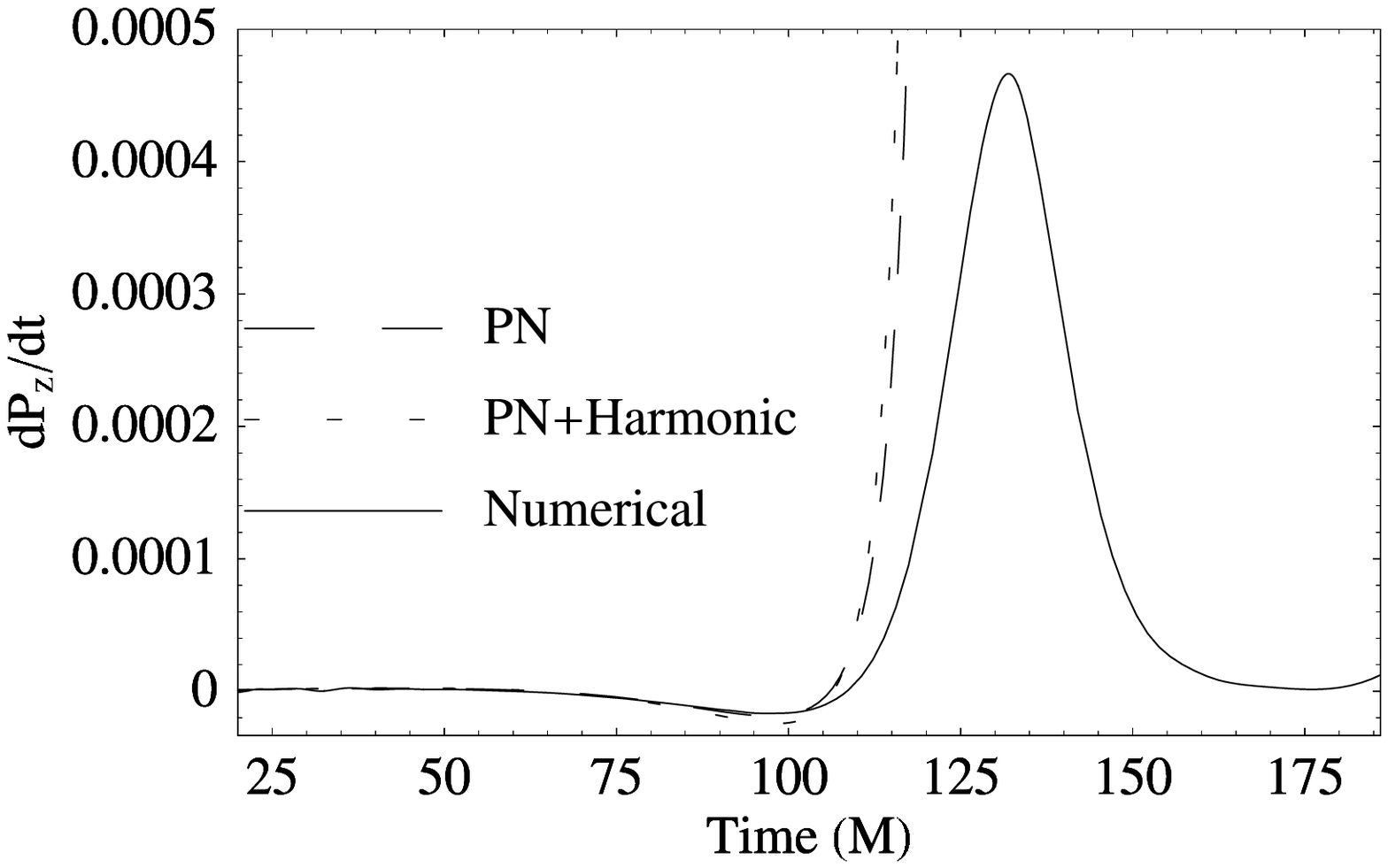}
\caption{\label{fig:HarmonicP} 
Comparison of the numerical $dP_z/dt$ with that predicted by 
Eq.~(\ref{eqn:KidderP}), using both the numerical puncture positions
and momenta in ADMTT coordinates, and those transformed to 
harmonic coordinates. At early times there is good qualitative agreement
between all three approaches. At late times (lower panel) the PN
and numerical values diverge, as discussed in Section~\ref{sec:PNcomparison}.
}
\end{figure}

When the particles are far apart and moving slowly, the
coordinates $\mathbf{x}_i$ and $\mathbf{X}_i$ will not differ much. In
Figure~\ref{fig:Harmonic}, which shows results from the D6 simulation, 
we show the separation between the punctures in the 
numerical coordinates as a function of time, and in the coordinates after the
transformation (\ref{eqn:ADMTTtoHarmonic}). The coordinates differ by less than 
10 \% up until about 15$M$
before merger. After that time we do not expect the coordinate transformation
(which is accurate up to only 2PN order) to be reliable. However, for most of
the evolution we see that the differences between the two coordinate choices
are not dramatic. A comparison of numerical and PN calculations of $dP_z/dt$ 
(as described in Section~\ref{sec:PNcomparison}) in Figure~\ref{fig:HarmonicP}
 also shows that the results
are similar at early times, before the PN result diverges. Note that when the 
puncture motion in ADMTT coordinates is used in the (harmonic) PN formula
(\ref{eqn:KidderP}), the curve is closer to the numerical result than when 
we use the puncture motion in harmonic coordinates. However, since this 
agreement occurs just before the time when the PN and numerical 
values seriously diverge, we do not take this agreement too seriously. 

The main conclusion of this analysis is that the difference in results between
using ADMTT and harmonic dynamical quantities in PN expressions is less 
than the uncertainty inherent in the PN expressions themselves. We therefore
continue to use the raw numerical data, in ADMTT coordinates, for most of
the analysis presented in this paper. 

\end{appendix}

\bibliography{refs}

\end{document}